\documentclass[APA,LATO1COL]{WileyNJD-v2}

\articletype{Special Issue Paper}%

\usepackage{multirow}
\usepackage{threeparttable}
\usepackage{subcaption} 

\begin{document}

\title{A compendium and evaluation of taxonomy quality attributes}

\author[1]{Michael Unterkalmsteiner*}

\author[1]{Waleed Abdeen}

\address[1]{\orgdiv{Software Engineering Research and Education Lab Sweden}, \orgname{Blekinge Institute of Technology}, \orgaddress{\country{Sweden}}}

\corres{*Michael Unterkalmsteiner, Valahallavägen 1, Karlskrona, 37141 Sweden. \email{michael.unterkalmsteiner@bth.se}}

\abstract[Abstract]{
\emph{Introduction:} Taxonomies capture knowledge about a particular domain in a succinct manner and establish a common understanding among peers. Researchers use taxonomies to convey information about a particular knowledge area or to support automation tasks, and practitioners use them to enable communication beyond organizational boundaries.
\emph{Aims:} Despite this important role of taxonomies in software engineering, their quality is seldom evaluated. Our aim is to identify and define taxonomy quality attributes that provide practical measurements, helping researchers and practitioners to compare taxonomies and choose the one most adequate for the task at hand.
\emph{Methods:} We reviewed 324 publications from software engineering and information systems research and synthesized, when provided, the definitions of quality attributes and measurements. We evaluated the usefulness of the measurements on six taxonomies from three domains.
\emph{Results:} We propose the definition of seven quality attributes and suggest internal and external measurements that can be used to assess a taxonomy's quality. For two measurements we provide implementations in Python. We found the measurements useful for deciding which taxonomy is best suited for a particular purpose.
\emph{Conclusion:} While there exist several guidelines for creating taxonomies, there is a lack of actionable criteria to compare taxonomies. In this paper, we fill this gap by synthesizing from a wealth of literature seven, non-overlapping taxonomy quality attributes and corresponding measurements. Future work encompasses their further evaluation of usefulness and empirical validation.
}


\maketitle

This is the peer reviewed version of the following article: Unterkalmsteiner, M., \& Adbeen, W. (2023). A compendium and evaluation of taxonomy quality attributes. Expert Systems, 40(1), e13098., which has been published in final form at \url{https://doi.org/10.1111/exsy.13098}. This article may be used for non-commercial purposes in accordance with Wiley Terms and Conditions for Use of Self-Archived Versions. This article may not be enhanced, enriched or otherwise transformed into a derivative work, without express permission from Wiley or by statutory rights under applicable legislation. Copyright notices must not be removed, obscured or modified. The article must be linked to Wiley’s version of record on Wiley Online Library and any embedding, framing or otherwise making available the article or pages thereof by third parties from platforms, services and websites other than Wiley Online Library must be prohibited.

\section{Introduction}
The development and use of taxonomies plays an important role in advancing and structuring knowledge in Software Engineering research~\citep{ralph_toward_2019} as well as Artificial Intelligence~\citep{ferrucci2013watson}. At the same time, taxonomies are used by researchers to solve software engineering problems (e.g. for classification of app reviews~\citep{panichella_how_2015}, trace recovery~\citep{li_ontology-based_2013}, technical debt classification~\citep{tom_exploration_2013}) and by practitioners to solve problems in the physical world (e.g. safety and human errors~\citep{wallace_beyond_2006}, business information exchange~\citep{debreceny_xbrl_2009}, classification of diseases~\citep{world_health_organization_classification_2021}).
Given the importance of taxonomies for both research and practice, one would assume that there exist commonly agreed quality attributes and evaluation approaches for taxonomies. However, while there exist several excellent guidelines on how to create taxonomies, they leave the definition of concrete quality attribute measurements for future research (e.g. \citet{nickerson_method_2013}) or focus only on a narrow set of quality attributes to assess the taxonomy's utility (e.g. \citet{usman_taxonomies_2017}). 

In our ongoing research, we are interested in using taxonomies for early requirements tracing~\citep{unterkalmsteiner_early_2020}. Since we have the choice of several domain-specific taxonomies for this task, we have to evaluate which taxonomy is likely the best fit in terms of structure and content. The goal of this paper is to establish a set of well-defined taxonomy quality attributes that can help us (and other researchers and practitioners in a similar situation) to make an informed decision when choosing taxonomies.

We define seven quality attributes (comprehensiveness, robustness, conciseness, extensibility, explanatory, mutual exclusiveness, and reliability) with corresponding internal and external measurements. We evaluate six domain specific taxonomies to demonstrate the usefulness of the proposed measurements.

The paper is structured as follows. Section~\ref{sec:rw} reviews related work. Section~\ref{sec:rm} provides an overview of our research method, followed by the literature survey of taxonomy quality attributes (Section~\ref{sec:results}). We present the quality attributes and measurements in Section~\ref{sec:compendium}, illustrate their application in Section~\ref{sec:comparison}, and discuss their usefulness and limitations in Section~\ref{sec:discussion}. The paper concludes in Section~\ref{sec:cfw}.

\section{Related work}\label{sec:rw}
Since software engineering relies heavily on the production and use of artifacts, the quality evaluation of engineering work products is a common research target. \citet{khosravi_quality_2004} reviewed seven quality models and synthesized a model to evaluate software design patterns. They identified 106 software quality characteristics, analyzed the relationship between these quality characteristics, identified quality metrics to measure some (25) of the most common quality characteristics, and specified a quality model for design patterns evaluation. Moreover, they evaluated 23 design patterns to assess the applicability of their approach. Their work is similar to ours, except our quality model goal is to evaluate taxonomies. \citet{basciani_tool-supported_2019} present a flexible approach to assess the quality of modeling artifacts using customizable quality models~\citep{basciani_customizable_2016}, and develop a tool to facilitate the assessment.
\citet{saavedra_software_2013} review nine quality models and propose a compendium of quality attributes for the evaluation of software requirements specifications. They map 23 quality attributes to evaluation techniques from the literature and show the interrelationship of different quality attributes.

%Ontology literature focuses on evaluation of learned ontologies? -> Argument to use why we need our paper?

%Modelling ontology evaluation and validation, 2006 -> highly relevant as very similar to our work, but focused on ontologies
% The main approaches for evaluation: structural, functional and usability-profiling.

%A survey on ontology evaluation methods, 2015 -> Categorize ontology evaluation approaches into four groups. Gold-standard approaches compare a learned ontology with a reference ontology. Corpus-based approaches evaluate to what extent a taxonomy covers a particular domain, represented by a corpus of domain-specific documents. Gold-standard and corpus-based approaches address accuracy, completeness and conciseness criteria of an ontology. Task-based approaches evaluate the improvement an ontology achieves for a particular task. Criteria-based approaches look at structural properties of the ontology. Task-based and criteria-based approaches address adaptability, clarity, computational efficiency and consistency of ontologies.    

%Monte Carlo study of taxonomy evaluation, 2010 -> not relevant

%TAXONOMY RESEARCH IN INFORMATION SYSTEMS: A SYSTEMATIC ASSESSMENT, 2019 -> not relevant

\citet{ralph_toward_2019} developed methodological guidelines for process theories and taxonomies in Software Engineering. He outlined three properties of good taxonomies: (1) the class structure enables the distinction between instances, (2) relevant properties of an instance can be inferred from class membership, and (3) the taxonomy fulfills the purpose for which it was designed. We argue that these are relevant, but not the only characteristics of good taxonomies. 
Two recent studies conducted reviews on the topic of taxonomy evaluation. \citet{usman_taxonomies_2017} surveyed the software engineering literature to understand how taxonomies are developed and used, and which subject matters are classified. Furthermore, they studied different proposals for taxonomy development and propose a synthesized method. Finally, they investigated how the utility of taxonomies is evaluated. A more recent study by \citet{szopinski_criteria_2020} reviewed taxonomies from the field of information systems, extracting taxonomy quality attributes, and proposed guidelines to assess the quality of taxonomies. Even though these studies' goal aligns with ours of determining the quality of taxonomies, they differ in two essential aspects. First, \citet{usman_taxonomies_2017} focused on the validation of taxonomies, which considers three perspectives: orthogonality of the taxonomy's dimensions and categories, the benchmarking against comparable taxonomies, and the demonstration of the taxonomies utility. 
From \citeauthor{szopinski_criteria_2020}'s \citeyearpar{szopinski_criteria_2020} work, we know that different studies have proposed 43 quality attributes to evaluate the created taxonomies. However, and secondly, we observed that Szopinski et al.'s work stops at describing \emph{what} has been evaluated, without answering in detail \emph{how} these identified quality attributes are measured.

To summarize, while the quality of other software engineering work products has been studied to a large degree, taxonomies have seen less attention. Those studies that investigated the quality attributes of taxonomies did not provide concrete means to assess them. The contribution of this paper is to fill this gap.  

\section{Research Methodology}\label{sec:rm}
The main goal of our research is to define a comprehensive set of quality attributes and associated measurements to evaluate taxonomies. To that end, we asked the following research questions:

\begin{enumerate}
    \item[\emph{RQ1}] How can the quality of a taxonomy be evaluated? 
    \begin{enumerate}
        \item[\emph{RQ1.1}] To what extent do the existing definitions of taxonomy quality attributes overlap?
        \item[\emph{RQ1.2}] What measurements exist to evaluate the quality of a taxonomy?
    \end{enumerate}
    \item[\emph{RQ2}] What insights can be gained by evaluating comparable taxonomies? 
\end{enumerate}

Note that in \emph{RQ1} we do not ask \emph{what} quality attributes exist for taxonomy evaluation. This question has already been answered with the reviews by \citet{szopinski_criteria_2020} and \citet{usman_taxonomies_2017}. To answer \emph{RQ1}, we re-analyzed the primary studies identified by \citet{szopinski_criteria_2020} and \citet{usman_taxonomies_2017} with the purpose of extracting and analyzing quality attribute definitions (answering \emph{RQ1.1}) and measurements for those attributes (answering \emph{RQ1.2}). We proceeded thereby as follows. 

The first author extracted, using the classification by \citet{szopinski_criteria_2020} as a key, verbatim definitions of quality attributes from the reviewed studies. If a study did only mention a quality attribute (with or without citation to a source), no definition was extracted. If a study described how the quality attribute was measured, we extracted that description verbatim or summarized it (in case a verbatim extraction was not feasible because there was no explicit concise description). In order to ensure that no data was overlooked, the second author performed an independent extraction. Then, the first author compared the extracted data and coded it according to the four categories shown in Table~\ref{tab:dac}. We discussed the disagreements in a meeting. Since we were interested to achieve high recall in the extraction, we focused on those disagreements (61) where the second author extracted more or different information than the first author ($A1 < A2$ and $A1 \neq A2$, column \citeauthor{szopinski_criteria_2020}). After the discussion, 47 data extractions led to no further changes as the diverging extractions could be explained by the fact that taxonomy quality attributes were often mentioned in different places in a paper (e.g. research methodology, results, discussion). The remaining 13 diverging extractions led to two types of changes: (a) a new distinct definition for a quality attribute was found (two instances), and (b) an already seen definition was found (11 instances) and added as a reference. 

\begin{table}[t]
    \caption{Data extraction comparison}
    \label{tab:dac}
    \centering
    \footnotesize
    \begin{tabular}{llll}
    \toprule
        Comparison & Description & \multicolumn{2}{c}{Frequency} \\
        & & \citeauthor{szopinski_criteria_2020} & \citeauthor{usman_taxonomies_2017} \\
    \midrule
        $A1 = A2$ & Both authors extracted essentially the same information (with varying amount of context) & 165 & 225 \\
        $A1 > A2$ & Author 1 extracted more information than Author 2 & 26 & 16 \\
        $A1 < A2$ & Author 2 extracted more information than Author 1 & 9 & 27 \\
        $A1 \neq A2$ & Author 1 extracted different information than Author 2 & 52 & 2 \\
        \midrule
        & Total number of extraction data points & 252 & 270 \\
    \bottomrule 
    \end{tabular}
\end{table}

Next, the first author sorted the quality attributes by number of found definitions and compared definitions within each attribute cluster to analyze similarities and differences. Similar definitions were grouped together, which led to the identification of differently named quality attributes that describe however the same concept. For example, the definitions of the quality attributes \emph{completeness}, \emph{expressiveness}, and \emph{generalizability} were all very similar to \emph{comprehensiveness}. On the other hand, we found also definitions of the same quality attribute that were contradicting or describing different concepts. We just considered quality attributes that defined a measurable concept. To illustrate this idea, we cite the definition of the quality attribute "compatibility with theories" by \citet{schaffer_assessing_2017}: "It [the taxonomy] is compatible with relevant theories. In the present case this implies that the taxonomy covers key concepts of coordination theory". The authors of the definition do not illustrate what these key concepts are and how the taxonomy covers them. Hence, we disregarded this quality attribute from our synthesis.  This process of constant comparison~\citep{dye2000constant}, whose results are presented in Section~\ref{sec:results}, led to the synthesis of seven quality attributes and measurements presented in Section~\ref{sec:compendium}. To validate the soundness of the synthesis, the second author reviewed the results that are traced back to the original definitions. We discussed in total 17 incongruences, of which 13 led to no change in the synthesis. The remaining changes were to add a paper as support for a definition (two instances), to reformulate the definition of a quality attribute (one instance), and adding an alias to a quality attribute (one instance). 

% How many discussions: 
% - of which agreed to change nothing: 13
% - changes: 1 (added paper to definition as support [Seyffarth, Conciseness), 1 (rewrote synthesis of an attribute [Chasin, feasibility] to reflect the definition more accurately, 1 (added paper to definition as support [Aksulu, Parsimoniousness], 1 (added new alias, unambiguousness, to reliability [Strasser]).

To evaluate the comprehensiveness of the synthesized quality attributes, the first author extracted evaluation descriptions from the 270 primary studies on taxonomies in Software Engineering, originally reviewed by~\citet{usman_taxonomies_2017}. The goal was to find new definitions of quality attributes that were not covered or contradicted the synthesized set from~\citep{szopinski_criteria_2020}. Again, to validate this process, the second author independently reviewed the set of 270 primary studies and mapped, where relevant and possible, the study to a synthesized quality attribute. The first author compared then the extracted data and coded it again according to the four categories shown in Table~\ref{tab:dac}. The results are reported in column \citeauthor{usman_taxonomies_2017} and led in total to 35 changes (33 new instances of already known quality attributes and two new alternatives names were found).

% Add paper to usefulness: 1
% Remove paper from applicability and add to reliability: 1
% Add paper to orthogonality: 2
% Added paper to usability: 1
% Added paper to applicability: 16
% Added paper to robustness: 1
% Added paper to completeness: 1
% Added paper to unambiguousness: 1
% Added paper to comprehensiveness: 5
% Added paper to reliability: 1
% Added paper to extensibility: 1
% Added paper to mutual exclusiveness: 2

% Added universality as synonym for comprehensiveness
% Added clarity as synonym for unambiguousness

To evaluate the usefulness of the proposed taxonomy quality attributes and their measurements, we applied them on six taxonomies from three domains (answering \emph{RQ2}). We purposefully chose two taxonomies for each domain. This allows us to discuss the differences between the taxonomies within a domain, using the quality attributes as a mean to steer the decision of choosing a taxonomy for a particular purpose. The data extraction for subjective measurements from the taxonomies was conducted by the second author and reviewed by the first author. The objective measurements were implemented by the first author.

To summarize, the research methodology followed in this study consisted of the following five steps:

\begin{enumerate}
    \item Identification and extraction of quality attribute definitions from Szopinski et al.'s primary studies.
    \item Synthesis of definitions and consolidation of quality attributes.
    \item Review of Usman et al.'s primary studies to identify new quality attributes or measurements.
    \item Identification and implementation of measurements, either from the analyzed primary studies or, if relevant, from the associated scientific literature.
    \item Evaluation of a set of taxonomies to assess the viability and usefulness of the defined quality attributes and measurements.
\end{enumerate}

The results of step (1), (2) and (3) are presented in Section~\ref{sec:results}, while the result of step (4), the compendium of taxonomy quality attributes and measurements, is presented in Section~\ref{sec:compendium}. The results of step (5) are reported in Sections~\ref{sec:comparison} and~\ref{sec:discussion}.  

\subsection{Validity threats}\label{sec:vt}
The main threat to the validity of this study is the reliability of the data extraction, interpretation and synthesis of the taxonomy quality attributes. In order to address this threat, two researchers were involved in executing the research method, performing the steps independently and recording any differences that were discussed in review meetings. Still, there is a moderate threat that researchers with a different background and without ongoing collaborations would make different decisions in the data extraction and synthesis. Therefore, we provide an extensive replication package (see Section~\ref{sec:replication}) that illustrates the taken decisions and can be used to verify our conclusions independently.

A secondary threat to validity is the generalizability of our results. First, the synthesized quality attributes might not be comprehensive, i.e. there might exist more relevant attributes than we identified. We mitigate this threat by not relying on a single data source (i.e. \citet{szopinski_criteria_2020}) and validate the synthesized list of quality attributes with the review by \citet{usman_taxonomies_2017}. Second, the suggested measurements to evaluate the quality attributes might be incomplete. This threat is moderate as we included in our synthesis only measurable quality attributes; still, the suggested measurements are likely only a subset of the possible ones and may not be applicable in every context. In particular, the proposed measurements assume that the taxonomy is hierarchical; measurements for faceted taxonomies are not provided but could be defined in future work. Third, the evaluation of the proposed quality attributes and measurements is conducted on a varied but still limited set (six taxonomies from three domains) of examples. However, the evaluation results do not indicate that the quality attributes and measurements are domain specific or do not scale with differently sized taxonomies.

\subsection{Replication package}\label{sec:replication}
Our research process resulted in several artifacts (data, trained models, source code, analysis results) that we share for the purposes of scientific transparency, replication, and reuse. The \emph{data} component contains the verbatim quality attribute definitions we found in the reviewed primary studies and data we produced during the validation of the data extraction and synthesis. We also provide the trained doc2vec \emph{model} that we used for the implementation in one of our proposed measurements, together with the source code for training eventual new models. The \emph{source code} component contains also the implementation of two  measurements. The \emph{analysis results} contain the results of applying the measurements on six taxonomies. The replication package is available as supplementary material~\citep{unterkalmsteiner_supp_2021}.

\subsection{Terminology}
\begin{figure}[t]
    \centering
    \includegraphics[width=0.6\columnwidth]{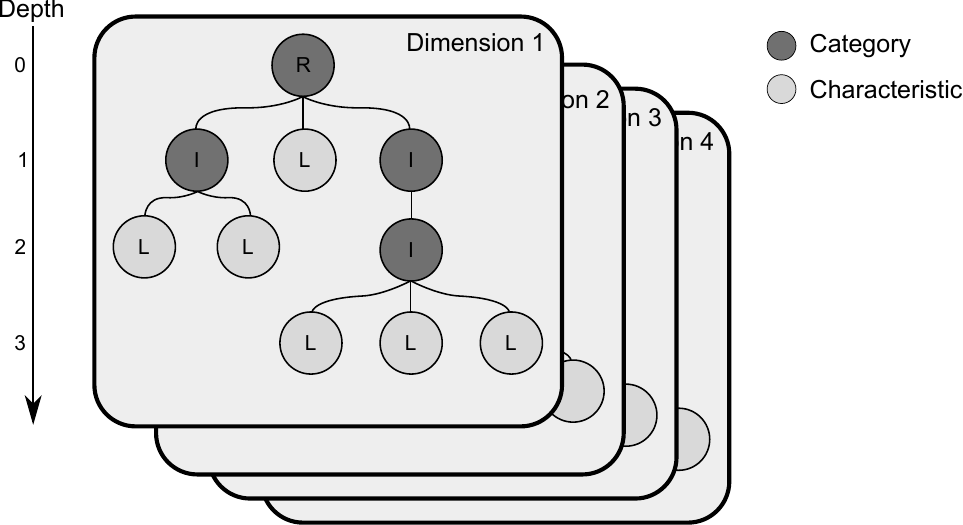} 
    \caption{Taxonomy represented as a hierarchy}
    \label{fig:taxcons}
\end{figure}
To be able to discuss taxonomy quality attributes and concrete measurements, we need to define the main constructs of a taxonomy. While there are different ways to structure a taxonomy, the most common one, and the one used in the taxonomies we are investigating, is a hierarchy. Figure~\ref{fig:taxcons} illustrates a hierarchy by means of a tree. A tree consists of nodes connected by edges, whereby the edges represent the parent-child relationships of the hierarchy. A tree has three types of nodes:
\begin{enumerate}
    \item (R)oot node: each tree has a single root node from which all other nodes descend.
    \item (I)ntermediate nodes: these are nodes between the root node and the leaf nodes. In the context of taxonomy constructs, we call these nodes \emph{categories}.
    \item (L)eaf nodes: these are nodes without children. In the context of taxonomy constructs, we call these nodes \emph{characteristics}. 
\end{enumerate}

Note that each node has exactly one parent (except the root node, which has no parent). Each node has a depth, which is the number of edges from the node to the tree's root node. The root node has a depth of 0. 

A taxonomy may consist of more than one tree in which case each tree represents one dimension of the taxonomy. The constructs that are relevant for the discussion of measurements of taxonomy quality attributes are the \emph{dimensions}, \emph{categories} and \emph{characteristics} of a taxonomy. Note however that these are not the only relevant aspects that are considered in taxonomy evaluation. The content of the taxonomy and the associated documentation on how the taxonomy was created and how to use the taxonomy are also relevant.

\section{Literature survey of taxonomy quality attributes}\label{sec:results}
In this section, we report on the identification and synthesis of quality attributes.

\subsection{Szopinski et al.}\label{sec:szopinksi}
\citet{szopinski_criteria_2020} reviewed 54 publications that developed and evaluated taxonomies for information systems research. From these publications, they identified 43 different taxonomy quality attributes. Table~\ref{tab:qa_is} lists the quality attributes for which we could identify distinct definitions and which alternative names emerged after our analysis, which is presented next. Note that for 17 quality attributes we could not identify any definition in the original publications. These quality attributes are listed and referenced to their origin at the end of this section.

\begin{table}[t]
    \caption{Quality attributes extracted from Szopinski et al.}
    \label{tab:qa_is}
    \centering
    \scriptsize
    \begin{tabular}{llp{10cm}}
    \toprule
         Attribute name & Distinct definitions & Alternative names \\
    \midrule
        usefulness & 5 & applicability\\ 
        comprehensiveness & 4 & collectively exhaustive, exhaustiveness, completeness, generalizability, inclusiveness \\ 
        conciseness & 3 & parsimoniousness, simplicity, no unnecessary dimensions, feasibility \\
        explanatory & 3 & - \\
        extensibility & 3 & modifiability \\
        robustness & 3 & distinctiveness \\
        applicability & 2 & usefulness \\
        collectively exhaustive & 2 & see comprehensiveness \\
        understandability & 1 & see reliability \\
        completeness & 1 & see comprehensiveness \\
        mutually exclusiveness & 1 & uniqueness \\
        reliability & 1 & understandability, repeatability, unambiguousness, construct validity, consistency \\
        consistency & 1 & see reliability \\
        construct validity & 1 & see reliability \\
        parsimoniousness & 1 & see conciseness \\
        repeatability & 1 & see reliability \\
        uniqueness & 1 & mutually exclusiveness \\
        distinctiveness & 1 & robustness \\
        exhaustiveness & 1 & see comprehensiveness \\
        feasibility & 1 & see conciseness \\
        generalizability & 1 & see comprehensiveness \\
        inclusiveness & 1 & see comprehensiveness \\
        modifiability & 1 & extensibility \\
        no unnecessary dimensions & 1 & see conciseness \\
        simplicity & 1 & see conciseness \\
    \bottomrule 
    \end{tabular}
\end{table}

\emph{Usefulness} was evaluated by a majority through the application of the taxonomy on sample data~\citep{puschel_whats_2016, gimpel_understanding_2018, strasser_delphi_2017, botha_towards_2018, stockli_capturing_2017, bock_towards_2017, siering_taxonomy_2017, williams_design_2008, gibbs_experience-based_2016, zrenner_data_2017}. In these cases, the operating definition of a useful taxonomy is whether it can be used to classify a known set of data. Other means to evaluate usefulness were to express usefulness through another set of quality attributes~\citep{aksulu_comprehensive_2010, diniz_taxonomy_2019, schneider_taxonomic_2014, oberlander_conceptualizing_2018, gibbs_experience-based_2016, botha_towards_2018, snow_developing_2016, thiebes_cancer_2017}, use of expert opinion~\citep{schaffer_assessing_2017, raza_transformation_2018, herzfeldt_developin_2012}, e.g. collected through interviews, and comparisons with other taxonomies~\citep{jarvinen_research_2000}. Some definitions were tautological, i.e. a taxonomy is useful if it is used~\citep{geiger_crowdsourcing_2012, snow_developing_2016}. An alternative name for usefulness is \emph{applicability}. A majority determines applicability by demonstrating the taxonomy on sample data~\citep{gregor_nature_2006, gimpel_understanding_2018, strasser_delphi_2017, bock_towards_2017, siering_taxonomy_2017, keller_reference_2014, conboy_agility_2009, fellmann_towards_2017}. Expert opinion through interviews is also used~\citep{tsatsou_towards_2010, werder_towards_2016}. 

A large set of studies determines \emph{comprehensiveness} by evaluating characteristics of the taxonomy creation method~\citep{prat_taxonomy_2015, schneider_taxonomic_2014, seyffarth_taxonomy_2017, snow_developing_2016, aksulu_comprehensive_2010, siering_taxonomy_2017, keller_reference_2014}. The selection of data sources, inclusion and exclusion criteria, and also the expertise of interviewees are often used to underpin that a taxonomy is comprehensive. Another interpretation of comprehensiveness is the taxonomy's ability to classify all known objects within a domain~\citep{nickerson_method_2013, strasser_delphi_2017, schaffer_assessing_2017, diniz_taxonomy_2019, botha_towards_2018, tonnissen_towards_2018, tilly_towards_2017, geiger_crowdsourcing_2012, snow_developing_2016, seyffarth_taxonomy_2017, prat_taxonomy_2015}. A closely related attribute is therefore \emph{generalizability}, that is, to classify objects that were not used in the process of taxonomy construction~\citep{gibbs_experience-based_2016}. One definition expresses comprehensiveness through another criterion, mutual exclusiveness~\citep{puschel_whats_2016}. Finally, a taxonomy is comprehensive if no new dimensions are discovered in future research~\citep{strode_dependency_2016}. \emph{Collectively exhaustive}~\citep{gregor_nature_2006, holler_defining_2017}, \emph{exhaustiveness}~\citep{ge_taxonomy_2018}, \emph{completeness}~\citep{schneider_taxonomic_2014, barenfanger_classifying_2016}, \emph{inclusiveness}~\citep{diniz_taxonomy_2019, botha_towards_2018} were all defined similarly to comprehensiveness, i.e. no items were found that could not be classified under the schema. However, another definition for collectively exhaustive is that classified items need to be assigned at least one characteristic under each dimension~\citep{diniz_taxonomy_2019}. 

A majority defined \emph{robustness} as the taxonomy's ability to differentiate objects of interest~\citep{nickerson_method_2013, schneider_taxonomic_2014, barenfanger_classifying_2016, schaffer_assessing_2017, snow_developing_2016, siering_taxonomy_2017, werder_towards_2016, labazova_hype_2018, tonnissen_towards_2018, strode_dependency_2016, tilly_towards_2017,aksulu_comprehensive_2010, seyffarth_taxonomy_2017}. This ability increases with the number of dimensions and characteristics, reduces however at the same time conciseness. Alternatively, robustness is defined as the lack of required changes to the structure of a taxonomy when new objects are classified~\citep{geiger_crowdsourcing_2012, chasin_peer--peer_2018}. One~\citep{fteimi_analysing_2018} publication defines a taxonomy as robust if it has a limited number of categories and subcategories, which is a definition of conciseness we found by other authors. 

A majority defined \emph{conciseness}~\citep{nickerson_method_2013, puschel_whats_2016, schneider_taxonomic_2014, barenfanger_classifying_2016, schaffer_assessing_2017, seyffarth_taxonomy_2017, snow_developing_2016, diniz_taxonomy_2019, botha_towards_2018, stockli_capturing_2017,siering_taxonomy_2017, werder_towards_2016, tonnissen_towards_2018, strode_dependency_2016} or \emph{simplicity}~\citep{prat_taxonomy_2015} through the taxonomy's number of dimensions and characteristics. \emph{Parsimoniousness} in the number of dimensions~\citep{jarvinen_research_2000} or categories~\citep{aksulu_comprehensive_2010} was also used to define conciseness. A concise taxonomy improves the ability to comprehend and apply the taxonomy. The recommendations for a concise taxonomy vary: two publications~\citep{schneider_taxonomic_2014, strode_dependency_2016} suggest to use \citeauthor{miller_magical_1956}'s \citeyearpar{miller_magical_1956} heuristic of $7\pm2$ items that a human can keep simultaneously in memory to determine the number of dimensions. Alternatively, one study determines conciseness by evaluating whether the taxonomy includes only meaningful dimensions~\citep{seyffarth_taxonomy_2017}, i.e. has dimensions that contain classified objects~\citep{labazova_hype_2018}. This is similar to the definition of \emph{no unnecessary categories}~\citep{gregor_nature_2006}. The \emph{feasibility} of a taxonomy is determined by whether data (objects) for a dimension do exist at all. \citet{chasin_peer--peer_2018} argue that non-feasible dimensions should be excluded, making the taxonomy more concise. Another definition~\citep{fteimi_analysing_2018}, that expresses rather comprehensiveness, states that in a concise taxonomy all found objects should be classified. 

A taxonomy is \emph{extensible}~\citep{nickerson_method_2013, prat_taxonomy_2015, schneider_taxonomic_2014, barenfanger_classifying_2016, schaffer_assessing_2017, fteimi_analysing_2018, diniz_taxonomy_2019, botha_towards_2018, siering_taxonomy_2017, werder_towards_2016, tonnissen_towards_2018, strode_dependency_2016, snow_developing_2016} or \emph{modifiable}~\citep{prat_taxonomy_2015} if it allows for the inclusion of additional dimensions and new characteristics when new types of objects are found. Variations to this definition are not common, e.g. the taxonomy allows only for adding new characteristics~\citep{seyffarth_taxonomy_2017} or combinations of existing characteristics~\citep{snow_developing_2016}. 

A taxonomy is \emph{explanatory} if it enables the user to locate an object in the taxonomy based on its characteristics or to deduce from the location of an object what characteristics it has~\citep{nickerson_method_2013, puschel_whats_2016, barenfanger_classifying_2016, schaffer_assessing_2017, fteimi_analysing_2018, diniz_taxonomy_2019, botha_towards_2018, werder_towards_2016, strode_dependency_2016, siering_taxonomy_2017, schneider_taxonomic_2014}, or more generally, provides useful explanations on current or future objects under study~\citep{snow_developing_2016}. In addition, \citet{keller_reference_2014} considered a taxonomy as explanatory when all characteristics are accompanied with an example. 

A \emph{mutually exclusive} taxonomy has no objects with two different characteristics in the same dimension~\citep{holler_defining_2017, diniz_taxonomy_2019, stockli_capturing_2017, labazova_hype_2018}. \emph{Uniqueness} was defined similarly as mutually exclusiveness~\citep{fteimi_analysing_2018, labazova_hype_2018}.

The classification through a taxonomy is \emph{reliable} if two or more independent classifiers agree on their classification decisions~\citep{puschel_whats_2016, oberlander_conceptualizing_2018, aksulu_comprehensive_2010, larsen_taxonomy_2003}. \emph{Understandability} was defined similarly as reliability, i.e. decision rules in a taxonomy are regarded as understandable if independent classifiers apply them consistently~\citep{gregor_nature_2006}. Furthermore, \emph{repeatability}~\citep{fteimi_analysing_2018}, \emph{construct validity}~\citep{oberlander_conceptualizing_2018, kupper_features_2014}, unambiguousness~\citep{strasser_delphi_2017}, and \emph{consistency}~\citep{aksulu_comprehensive_2010} were defined similarly as reliability.

For 17 quality attributes, no definitions were provided in the respective studies: utility~\citep{almufareh_taxonomy_2018, king_empirical_1999, labazova_hype_2018, alrige_toward_2015}, efficiency~\citep{almufareh_taxonomy_2018, botha_towards_2018, beevi_data_2015}, stability~\citep{bapna_user_2004, ge_taxonomy_2018}, sufficiency~\citep{gao_rethinking_2018}, effectiveness~\citep{chasin_peer--peer_2018, herterich_understanding_2015}, adequateness~\citep{schaffer_assessing_2017}, compatibility with theories~\citep{schaffer_assessing_2017}, purposefulness~\citep{strasser_delphi_2017}, usability~\citep{diniz_taxonomy_2019}, descriptiveness~\citep{raza_transformation_2018}, 
versatileness~\citep{raza_transformation_2018}, sufficiently detailedness~\citep{tonnissen_towards_2018}, appropriate wording~\citep{herterich_understanding_2015}, relevance~\citep{herterich_understanding_2015}, real world fidelity~\citep{johnk_how_0217}, face validity~\citep{king_empirical_1999}, suitability~\citep{cledou_taxonomy_2018}.

\subsection{Usman et al.}\label{sec:usman}
\begin{table}[t]
    \caption{Quality attributes extracted from Usman et al.}
    \label{tab:qa_se}
    \centering
    \scriptsize
    \begin{tabular}{lll}
    \toprule
         Attribute name & Alternative names & Frequency  \\ 
    \midrule
        applicability & usefulness, effectiveness, usability & 41 \\
        comprehensiveness & completeness, expressiveness, generalizability, universality & 21 \\
        reliability & objectivity & 10 \\
        usefulness & see applicability & 9 \\
        orthogonality & mutual exclusiveness & 5 \\
        conciseness & no unnecessary axes, no unnecessary categories & 3 \\
        usability & see applicability & 3 \\
        completeness & see comprehensiveness & 3 \\
        mutual exclusiveness & orthogonality & 3 \\
        robustness & distinguishability & 3 \\
        extensibility & - & 3 \\
        unambiguousness & clarity & 2 \\
        no unnecessary axes & see conciseness & 1 \\
        objectivity & reliability & 1 \\
        granularity & see conciseness & 1 \\
        expressiveness & see comprehensiveness & 1 \\
        effectiveness & see applicability & 1 \\
        distinguishability & robustness & 1 \\
        explanatory & - & 1 \\
        generalizability & see comprehensiveness & 1 \\
        no unnecessary categories & see conciseness & 1 \\
        clarity & unambiguousness & 1 \\
        universality & see comprehensiveness & 1 \\
    \bottomrule 
    \end{tabular}
\end{table}

\citet{usman_taxonomies_2017} reviewed 270 primary studies from software engineering that develop and propose a taxonomy. While Usman et al. asked in their review whether the taxonomies are validated, they did not investigate whether the taxonomies were created by considering certain quality attributes. Therefore, we re-analyzed the primary studies identified by Usman et al., extracting statements that indicate whether certain quality attributes were considered. Table~\ref{tab:qa_se} illustrates the results of our analysis.

\citet{usman_taxonomies_2017} reported that 66\% of the taxonomies were validated by either illustration, case study, experiment, expert opinion or survey. The goal of the validation is to demonstrate utility, which is equivalent to the quality attributes \emph{usefulness} and \emph{applicability} identified by Szopinski et al. \emph{Applicability} was also the most frequently evaluated quality attribute, followed by \emph{comprehensiveness}, \emph{reliability}, \emph{conciseness}, \emph{robustness}, \emph{extensibility} and \emph{orthogonality}. 
Overall, we could identify taxonomy quality attributes in only 56 out of 270 papers from the Software Engineering literature. All attributes correspond to the ones identified in the Information Systems literature, i.e. we did not find any new taxonomy quality attribute in the Software Engineering literature. 

\subsection{Synthesis of quality attributes}
The main goal of extracting and analyzing quality attribute definitions was to reduce the multiplicity of names for the same attribute and determine a minimal, well defined set of measurable quality attributes. In Section~\ref{sec:szopinksi} we have grouped quality attributes according to their definitions from the Information Systems literature. The review of taxonomies published in the Software Engineering community (Section~\ref{sec:usman}) did not reveal any new quality attributes.

Usefulness and applicability were the most commonly reported quality attributes to assess taxonomies. The evaluation usually consists of applying the taxonomy on a sample data set and determining how "well" the taxonomy can classify the data. This is aligned with \citeauthor{ralph_toward_2019}'s \citeyearpar{ralph_toward_2019} third criterion for a good taxonomy, i.e. the taxonomy fulfills the purpose for which it was designed. Alternatively, the taxonomy is assessed by a number of domain experts with respect to its usefulness or applicability. While these quality attributes are the most common, they are also subjective and do not provide any insight on the nature of the taxonomy. None of the studies suggested \emph{how} applicability or usefulness can be assessed in an objective manner, except in a few cases where other attributes, such as conciseness or reliability and their respective measures were used a proxy. We argue therefore that for the assessment of a taxonomy's quality, applicability and usefulness are merely placeholders for the more essential attributes in our proposed compendium of taxonomy quality attributes, \emph{comprehensiveness}, \emph{robustness}, \emph{conciseness}, \emph{extensibility}, \emph{explanatory}, \emph{mutual exclusiveness} and \emph{reliability}, which we introduce next.

\section{Taxonomy quality attributes and measurements}\label{sec:compendium}
We formulate taxonomy quality attributes and measurements by utilizing principles from measurement theory, in particular borrowing from \citeauthor{fenton_software_2014}'s~\citeyearpar{fenton_software_2014} approach to software metrics. We think that these principles, even though formulated in the context of software engineering practice, are equally applicable to the evaluation of taxonomies. 

The entities that can be potentially evaluated are (i) the process of creating the taxonomy, (ii) the products of this process, one of which is the taxonomy itself, but also other artifacts as user documentation, and (iii) the resources needed to create or to use/operate the taxonomy. We have collected and synthesized the quality attributes in Section~\ref{sec:results}, and define for each four characteristics. The \emph{description} provides a consolidated definition, based on the common definitions we found in the literature. \emph{Aliases} provide common other names under which the quality attribute might be known. We decided to add this characteristic to ease the identification of quality attributes and also make their equivalence relationship explicit. The \emph{nature} of the attribute is either internal or external. An internal attribute can be measured solely by observing the product, while an external attribute can only be observed in relation of the product and its behavior in a certain environment. Finally, the \emph{measurement} defines concrete means which can be used for the evaluation of the quality attribute. Several measurements we propose draw inspiration from the work we reviewed in Section~\ref{sec:results} or are based on our experience with analyzing textual data and taxonomies. Each measurement is further qualified by its \emph{nature}, \emph{perspective}, \emph{mapping} and \emph{scale}.

Similar to an attribute, a measurement can be internal or external. An internal measurement operates on data that stems purely from the product itself, while an external measurement operates on data that is collected from observing the behavior of the product in a particular environment. Therefore, an internal measurement is typically less costly to implement than an external measurement and can be collected even during the development of the product~\citep{fenton_software_2014}.  
This dichotomy of internal/external attributes and measurements opens the interesting question on whether an external attribute can be evaluated by an internal measurement\footnote{The reverse, evaluating an internal attribute with an external measurement, is conceivable but makes little practical sense in the presence of internal measurements, given the generally higher cost of collecting external measurements.}. In Software Engineering, a substantial amount of research has invested in determining whether internal measurements can predict external attributes~\citep{fenton_software_2014}. Unfortunately, it is not easy to establish a causal relationship between internal measurements and external attributes, and requires extensive empirical experimentation~\citep{fenton_software_2014}.

The perspective of a measurement can either be subjective or objective~\citep{fenton_software_2014}. A subjective measurement depends on the qualities (knowledge, skill, experience, bias) of the person that measures. Objective measures are more precise, in the sense that repeated measuring results in the same outcome and do not depend on personal judgment. Nevertheless, subjective measures can be useful as long as their imprecision is acknowledged and considered in the decision process~\citep{fenton_software_2014}.

The mapping of an attribute to a measurement can be direct or derived~\citep{fenton_software_2014}. In a direct measurement, no other attributes are involved, whereas in a derived measurement multiple attributes are combined.

The scale of a measurement defines which operations are allowed on the symbols that represent the measure that has been collected from empirical observations~\citep{fenton_software_2014}. A commonly used differentiation is the nominal, ordinal, interval, ratio and absolute scale~\citep{fenton_software_2014}. 

Table~\ref{tab:qam} provides an overview of the attributes and measurements in terms of the just presented measurement principles. For all attributes, except for reliability, we have identified internal as well as external measurements. Our paper does not attempt to establish a causal relationship between the proposed internal measurements and the corresponding external attributes. Internal measurements can still be useful even if they are only based on intuition~\cite{fenton_software_2014}, in particular if the purpose is to provide a relative comparison between two or more products.

\begin{table}[t]
    \caption{Taxonomy quality attributes and measurements}
    \label{tab:qam}
    \centering
    \scriptsize
    \begin{tabular}{lllllll}
    \toprule
         \multicolumn{2}{c}{Attribute} & \multicolumn{5}{c}{Measurement} \\ 
    \midrule
        Name & Nature & Name & Nature & Perspective & Mapping & Scale \\
    \midrule
        \multirow{2}{*}{Comprehensiveness} & \multirow{2}{*}{External} & Design process quality & Internal & Subjective & Direct & Nominal \\
        & & Count of unclassified objects & External & Subjective & Direct & Absolute \\
    \midrule
        \multirow{3}{*}{Robustness} & \multirow{3}{*}{External} & Count of taxonomy constructs & Internal & Objective & Direct & Absolute \\
        & & Semantic proximity and distance & Internal & Objective & Derived & Ratio \\
        & & Count of misclassified objects & External & Subjective & Direct & Absolute \\
    \midrule
        \multirow{3}{*}{Conciseness} & \multirow{3}{*}{External} & Human memory heuristic & Internal & Objective & Direct & Absolute \\
        & & Amount and depth of constructs & Internal & Objective & Derived & Ratio \\
        & & Count of unused constructs & External & Subjective & Direct & Absolute \\
    \midrule
        \multirow{2}{*}{Extensibility} & \multirow{2}{*}{Internal} & Documented extension points & Internal & Subjective & Direct & Nominal \\
        & & Rate of change & Internal & Objective & Derived & Ratio \\
    \midrule
        \multirow{2}{*}{Explanatory} & \multirow{2}{*}{External} & Location support & Internal & Subjective & Direct & Nominal \\
        & & Orientation efficiency and effectiveness & External & Objective & Derived & Ratio \\
    \midrule
        \multirow{2}{*}{Mutual exclusiveness} & \multirow{2}{*}{External} & Classification constraint & Internal & Subjective & Direct & Nominal \\
        & & Ambiguous objects & External & Objective & Derived & Ratio \\
    \midrule
        \multirow{2}{*}{Reliability} & \multirow{2}{*}{External} & Inter-rater reliability & External & Objective & Derived & Ratio \\
        & & Intra-rater reliability & External & Objective & Derived & Ratio \\    
    \bottomrule 
    \end{tabular}
\end{table}

%% Scale assignments based on definitions by:
%% Software Metrics: A Rigorous and Practical Approach (Fenton and Bieman)
%% On the theory of scales of measurement (Stevens)
%% http://media.acc.qcc.cuny.edu/faculty/volchok/Measurement_Volchok/Measurement_Volchok5.html

\subsection{Comprehensiveness}
\paragraph{Definition} Comprehensiveness is the taxonomy's ability to classify all known objects for the domain it was developed for. The attribute is an external quality as it is observed as part of the taxonomy's application.

\noindent\emph{Aliases:} collectively exhaustive, exhaustiveness, completeness, generalizability, inclusiveness, expressiveness, universality

\subsubsection{Design process quality}
In software engineering, the quality of the development process is often associated with the quality of the resulting software product~\cite{cugola1998software, fuggetta2000software}. 
The design process of a taxonomy is also a complex undertaking that, while guidelines exist (e.g.~\citet{nickerson_method_2013, usman_taxonomies_2017}), is difficult to assess without having taken part in the process. Nevertheless, the documentation that accompanies a taxonomy might contain pieces of evidence on the quality of the process, and therefore on the comprehensiveness of the taxonomy:
\begin{itemize}
    \item The data sources that were considered when identifying dimensions, categories and characteristics of the taxonomy. These sources may include standards, scientific literature, or even a sample of the to be classified objects. A higher diversity in data sources may indicate higher comprehensiveness. \emph{Question: Does the taxonomy originate from a diverse set of data sources?} 
    \item The diversity of the people or organizations involved in the creation of the taxonomy. A higher diversity in cultural and educational background may indicate higher taxonomy comprehensiveness. \emph{Question: Do the taxonomy creators have a diverse cultural and educational background?}
    \item The creators of the taxonomy might employ different means to analyze the data sources. Different data analysis approaches leading to the same outcome (i.e. taxonomy) indicate high comprehensiveness. \emph{Question: Were diverse data analysis methods used to create the taxonomy?} 
    \item The independent evaluation by external experts provides a check against biases and limitations in culture and education of the taxonomy creators. \emph{Question: Has the taxonomy been evaluated by external experts?}
\end{itemize}
Each of the above questions can be answered with a yes/no answer and are therefore a direct measure on a nominal scale. We do not count positive and negative answers (hence nominal scale) as the list can likely be arbitrarily reduced or extended, i.e. there is no "true" number of questions that is correct. We classified the measurement as subjective since the fidelity of the answers depends on the evaluators ability to identify and interpret information pertaining to the questions. The measure is internal as it is based on observation of the taxonomy's accompanying documentation and not its behavior.

\subsubsection{Count of unclassified objects}\label{sec:cuo}
When the taxonomy is applied, for its intended purpose and within scope, one can count the objects that could not be classified. This measure provides evidence for the lack of comprehensiveness of a taxonomy, and is an external and direct measure. It is also subjective since it depends on the users ability to correctly apply the taxonomy. The scale is absolute since we are counting objects and have a natural zero point, i.e. the empty set of no unclassified objects. Any object that cannot be classified can then in turn be analyzed in order to extend the taxonomy.

\subsection{Robustness}
\paragraph{Definition} Robustness is the taxonomy's ability to differentiate objects of interest. This ability is determined by the degree to which a taxonomy's categories and characteristics represent distinct concepts (orthogonality). Furthermore, the more closely related the characteristics in a category are (cohesiveness), the easier it is to classify an object. This definition is aligned with \citeauthor{ralph_toward_2019}'s \citeyearpar{ralph_toward_2019} first criterion for a good taxonomy, i.e. that the classification structure enables the distinction between instances. The attribute is an external quality as it is observed as part of the taxonomy's application.

\noindent\emph{Aliases:} distinctiveness, distinguishability

\subsubsection{Count of taxonomy constructs}
\citet{nickerson_method_2013} suggest to measure robustness by counting the number of dimensions, categories and characteristics of a taxonomy, which is an internal, objective, direct measure on the absolute scale. They argue that a small number may not adequately be able to differentiate between objects of interest. By proposing this measurement, \citet{nickerson_method_2013} introduce thereby a trade-off between robustness and conciseness. Conciseness, which can also be measured by counting the number of dimensions, categories and characteristics of a taxonomy, would decrease when robustness increases.

\begin{figure}[t]
    \centering
    \includegraphics[width=0.5\columnwidth]{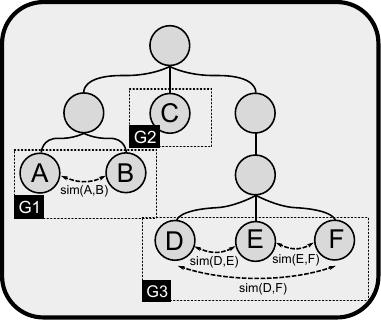} 
    \caption{Robustness example}
    \label{fig:robex}
\end{figure}

\subsubsection{Semantic proximity and distance of taxonomy constructs}\label{sec:robmes}
In order to avoid this trade-off, we propose an alternative measurement for taxonomy robustness. Instead of counting taxonomy constructs, we analyze the orthogonality and cohesiveness of the taxonomy's \emph{content}. A cohesive category contains characteristics that are in semantic proximity, while orthogonal categories contain characteristics that are in semantic distance, i.e. not related to each other. We use word embeddings~\citep{mikolov_distributed_2013} to model semantic similarity of characteristics, based on their distributional properties that we have learned from a large text corpus\footnote{All English Wikipedia articles.}. 
We illustrate the overall idea in Figure~\ref{fig:robex} and provide an intuition of the algorithm. First we identify all groups of leaf nodes in the taxonomy tree, indicated as G1, G2 and G3 in Figure~\ref{fig:robex}. For each group, we determine all possible pairs and calculate the pairs' similarity, i.e. G1: $sim(AB)$, G2: N/A, G3: $sim(DE)$, $sim(DF)$, $sim(EF)$. 
Each group (except G2, since it contains no pairs) contains a pair of characteristics with the smallest similarity, determining the cohesiveness of the group. For G1, this is trivial since there is only one pair (AB). For G3 it is $minimum(sim(DE), sim(EF), sim(DF))$. In order to quantify the orthogonality of a group, we introduce the concept of an \emph{intruder} characteristic. For each characteristic in a group, we calculate their similarity with every other characteristic \emph{outside} that group. If that similarity is higher than the minimum pairwise similarity calculated earlier, we increase the \emph{intruder} counter for that group by one. We aggregate the occurrence of intruder characteristics in the overall taxonomy as follows:

\begin{equation}
\label{eq:robeq}
    R(T) = \frac{\sum_{i=1}^{ngroups}(1 - \frac{n_{ic}}{n_{gc} * (n_{ac} - n_{gc})})}{ngroups}
\end{equation}

where $n_{ic}$ is the number of intruder characteristics, $n_{gc}$ is the number of group characteristics, and $n_{ac}$ is the number of all characteristics in the taxonomy. The range for robustness is $(0\ldots1]$ where 1 signifies that no intruder characteristics were found. The measure is internal and objective, as well as derived as we construct it from measuring similarity of taxonomy constructs and introduce the concept of intruder characteristics (an abstraction that does not exist, as such, in the taxonomy). Furthermore, the measure is on a ratio scale as we essentially count properties (intruder characteristics) but scale the count in equation~\ref{eq:robeq} to make the results comparable between different taxonomies.

\subsubsection{Count of misclassified objects}
When the taxonomy is applied, for its intended purpose and within scope, one can count the misclassified objects. If two or more objects are classified in the same category, even though they do not share characteristics that justify such a classification, the taxonomy is not robust. The measure is external, direct, subjective (since it depends on the users ability to apply the taxonomy), and on an absolute scale. 

\subsection{Conciseness}

\paragraph{Definition} Conciseness is the taxonomy's ability to classify objects of interest with the least possible amount of dimensions, categories and characteristics. The attribute is an external quality as it is observed as part of the taxonomy's application.

\noindent\emph{Aliases:} parsimoniousness, simplicity, no unnecessary dimensions/categories, feasibility

\subsubsection{Human memory heuristic}
The human working memory is limited to hold a few items at a time~\citep{miller_magical_1956}. We can construct a measure that is based on the number of taxonomy constructs a user needs to hold in memory to perform the classification task. At every decision point, we count the number of items the user needs to keep in memory and increment the counter $M_t$ if the threshold $t$ (for example 5) was surpassed. The decision points are the following:
\begin{itemize}
    \item Choice of dimension
    \item Choice of category at level $[1...n - 1]$. 
    \item Choice of characteristic at level $n$.
\end{itemize}
The higher $M_t$, the less concise is the taxonomy. The measure is internal, objective, direct, and on an absolute scale. 
The application of this measure is only meaningful if the taxonomy is used by humans since computational devices, which may use taxonomies e.g. for information retrieval purposes, are not limited to such a small amount of memory slots.

\subsubsection{Amount and depth of constructs}
The measurement based on human memory assumes that decisions are made locally and one after the other. This is likely a too simplistic view how taxonomy users make classification decisions. Assuming that decisions can be faulty, the user needs to backtrack the taxonomy tree while keeping the evidence for the wrong decision in memory and find a better choice. The measurement that we describe next assumes that wrong decisions and backtracking are the norm rather the exception. \citet{prat_taxonomy_2015} propose the following measure for taxonomy conciseness:

\begin{equation}
    C(T) = \frac{1}{1 + \ln{(\sum_{i=1}^{ncat}(\frac{1}{d(Cat_i)}) + \sum_{j=1}^{nchar}(\frac{1}{d(Char_j)}) - 1)}}
\end{equation}

The measurement depends on the number of categories ($ncat$) and characteristics ($nchar)$ as well as their depth $d$, and is hence a derived measure. The intuition behind considering depth as an influencing factor is that adding a category or characteristic at a lower depth (closer to the root) reduces conciseness more than adding a category or characteristic at a higher depth. The range for conciseness is $(0\ldots1]$ where the simplest possible taxonomy (1 category and 2 characteristics) evaluates to $C(T)=1$. The measure is internal, objective, and on a ratio scale (there is a natural zero, even though it can only be approached with an infinite depth of categories and characteristics).

\subsubsection{Count of unused constructs}
When the taxonomy is applied, for its intended purpose and within scope, one can count the categories and characteristics that are not used. A meaningful measurement is however only possible if a representative sample of the object population is drawn and classified. Defining such a sample is challenging. One could randomly sample and classify objects, and plot the frequency of newly used categories or characteristics. One would expect that with the increase of classified objects, fewer never used categories and characteristics are identified. When this level of saturation is reached, we can count the unused constructs. This measurement is external, direct, subjective (since it depends on the users ability to apply the taxonomy), and on an absolute scale.

\subsection{Extensibility}\label{sec:extensibility}
\paragraph{Definition} Extensibility is the taxonomy's ability to allow for changes in its structure, i.e. adding, modifying or deleting dimensions, categories or characteristics. The attribute is internal as it can be observed from the taxonomy structure.

\noindent\emph{Aliases:} Modifiability

\subsubsection{Documented extension points}
The documentation that accompanies a taxonomy might contain a change process which describes how the taxonomy can be extended. We can compare taxonomies' extensibility based on: 
\begin{itemize}
    \item whether the change process covers the following constructs: dimensions, categories, characteristics
    \item whether the change process covers the following change types: addition, modification, deletion
    \item who has the mandate to change the taxonomy
    \item how the taxonomy can be customized for use in a particular context
\end{itemize}

When extracting information on the above items from the documentation it is necessary to distinguish between the case where there is indeed no change process foreseen and a lack of information on the change process. Since the ability to identify documentation and extract information depends on the abilities of the evaluator, the measure is subjective. Furthermore, it is internal, direct and on a nominal scale. 

\subsubsection{Rate of Change}
An additional indicator of a taxonomy's extensibility is the rate how often it changes. Note however that this internal measurement alone can be misleading as mature taxonomies are likely less often changed, i.e. are stable, as compared to newly created taxonomies. We define rate of change as:

\begin{equation}
    RoC_{now} = \frac{n}{t_{now} - t_{initial\_release}}
\end{equation}

where $n$ is the number of changes (which can be either the actual changes if they are published, or the number of releases of the taxonomy), and $t_{now} - t_{initial\_release}$ is the number of days between the initial release of the taxonomy and now. The range for rate of change is (0\ldots1]. A higher value signifies that the taxonomy has changed more frequently, indicating extensibility but also lower stability. The measurement is objective, derived (since it depends on the number of releases and the times of releases), and on a ratio scale.

\subsection{Explanatory}
\paragraph{Definition} A taxonomy is explanatory if it enables the user to locate an object in the taxonomy based on its characteristics, or to deduce from the location of an object what characteristics it has. This definition is aligned with \citeauthor{ralph_toward_2019}'s \citeyearpar{ralph_toward_2019} second criterion for a good taxonomy, i.e. that relevant properties of an instance can be inferred from class membership. The attribute is an external quality as it is observed as part of the taxonomy's application. 

\noindent\emph{Aliases:} None

\subsubsection{Location support}
This is a difficult to evaluate attribute since, while it is relevant for the usability of the taxonomy, the definition does not provide many indications how the location support can be realized. We use therefore two examples from taxonomies we have reviewed and formulate questions that can be extended to a longer list when such examples appear in future work.
\begin{enumerate}
    \item \emph{Has the taxonomy any other structuring elements besides dimensions?} \citet{puschel_whats_2016} cluster the dimensions in their taxonomy into layers, claiming that this increases the taxonomy's explanatory strength.
    \item \emph{Does the taxonomy provide any support for choosing a particular dimension or category?} \citet{strode_dependency_2016} formulates questions as classification decision rules that support the taxonomy user in their choice.
\end{enumerate}

We classified the measurement as subjective as the fidelity of the answers depends on the evaluators ability to identify the information. Furthermore, the measure is internal, direct and on a nominal scale.

\subsubsection{Orientation efficiency and effectiveness}
With these external measurements, we can assess how efficient (in terms of decision timeliness) and effective (in terms of decision correctness) the taxonomy is when a user needs to determine the location of an object in the taxonomy. The measurements require a controlled environment in which such decisions can be observed, recorded and analyzed. For example, an experiment could be set up in which the participants have two tasks: 1) inspect an object and determine its location (dimensions, categories) based on its characteristics, and 2) be given a dimension, category and object, and deduce the objects characteristics. For both tasks, we could measure the time used and the number of correct decisions, normalized by the size of the taxonomy. Assuming there exists a correct answer for the decisions, the measurements are objective. The measurements are furthermore derived (due to the normalization by taxonomy size) and on a ratio scale.  

\subsection{Mutual exclusiveness}
\paragraph{Definition} Mutual exclusiveness is the taxonomy's ability to identify an object uniquely, i.e. that no object exists in the same dimension under different categories. The attribute is an external quality as it is observed as part of the taxonomy's application. 

\noindent\emph{Aliases:} Uniqueness, orthogonality

\subsubsection{Classification constraint}
The documentation that accompanies a taxonomy might contain explicit instructions on how to proceed in the case of a mutual exclusiveness violation. For example, the repeated violation of a classification constraint could lead to a taxonomy change request. We can compare taxonomies on whether a mutual exclusiveness classification constraint exists or not. This measurement is internal, subjective, direct and on a nominal scale. 

\subsubsection{Ambiguous objects}
If an object is classified in different categories, the classification is ambiguous. The measurements suggested for reliability can be used to assess this attribute.

\subsection{Reliability}\label{sec:reliability}
\paragraph{Definition} Reliability is the taxonomy's ability to support consistent classification decisions. The attribute is an external quality as it is observed as part of the taxonomy's application. 

\noindent\emph{Aliases:} Understandability, repeatability, unambiguousness, construct validity, consistency, objectivity

\subsubsection{Inter-rater reliability}
Inter-rater reliability refers to the consistency of how \emph{two or more raters} classify the same object. This measures whether the classification rules are interpreted in the same way by different people and, given the same object, lead to the same classification.
There exist different means to assess inter-rater reliability~\citep{gwet_handbook_2014}, for example Kappa statistics (Cohen, Fleiss), correlation coefficients (Pearson, Kendall, Spearman), intra-class correlation coefficient and Krippendorff's alpha. These measures are external, objective, derived and on a ratio scale.

\subsubsection{Intra-rater reliability}
Intra-rater reliability refers to the consistency of \emph{one} rater classifying the same or a similar object consistently at different occasions. Again, this measures whether the classification rules are interpreted the same way, this time however by the same rater seeing similar objects at different times. The same metrics as for inter-reliability can be used.

\section{A comparison of domain specific taxonomies}\label{sec:comparison}

In order to evaluate the usefulness of the proposed quality attributes and measures, and to illustrate their application, we chose six taxonomies from three domains. Our choice is based on our previous experience working with these taxonomies in research projects (construction domain) and the availability of two independent taxonomies on the same subject (cyber security and economic activities). 

To be able to choose a particular taxonomy, one must give either weights or priorities to the quality attributes. In our opinion, it is more intuitive to prioritize quality attributes as the order can be motivated by the particular classification use case in which the taxonomy is applied. We argue that for classification, the most important attributes are robustness, comprehensiveness and mutual exclusiveness. These determine how useful a taxonomy is for its main purpose. However, depending on whether the taxonomy is used by humans or by an automated classifier, conciseness and explanatory power could be equally important. A human classifier probably finds a concise and explanatory taxonomy more useful. An automated classifier is less likely dependent on those attributes. Finally, for the purpose of a classification task, extensibility seems to be the least relevant, except if the classification is expected to require modifications in the taxonomy. A formal alternative or complement to prioritization can be the Goal-Question-Metric paradigm~\citep{basili_goal_1994, fenton_software_2014} that would allow evaluators to choose metrics based on the goals formulated for taxonomy application.

In this evaluation, we assume the classification task is to be automated. Hence, we prioritize the quality attributes as follows: Robustness~$>$ Comprehensiveness~$>$ Mutual exclusiveness~$>$ Conciseness~$>$ Explanatory~$>$ Extensibility.
In the remainder of this section, we illustrate the application and show the usefulness of the proposed quality attributes to choose between two alternative taxonomies. We order the discussion of quality attributes according to the priority we defined here, and end each domain example with a motivated decision.

Table~\ref{tab:eval} provides a summary of the evaluation result. Please note that the entries for subjective measurements (comprehensiveness, extensibility, explanatory and mutual exclusiveness) are based on data extracted from the documentation of the taxonomies. For example, the answer to the question "Do the taxonomy creators have a diverse background?" was answered with "yes" and "no" for Omniclass and Mahaini respectively. This is based on the following data/meta-data:
\begin{itemize}
    \item Uniclass: The OCCS Development Committee acknowledges the contributions of over one hundred individuals over the last five years, many representing the following organizations, who assisted with the development of OmniClass. A list of organizations can be found online https://www.csiresources.org/standards/omniclass/standards-omniclass-contributors).
    \item Mahaini: The paper in which the taxonomy was proposed was written by three authors, two working at the same university.
\end{itemize}

The data we collected for the remaining taxonomies and questions is available in the supplementary material~\citep{unterkalmsteiner_supp_2021}.

It is worth to note that we chose only internal measurements for the comparison as implementing the external measures was prohibitively expensive, both in terms of effort and in terms of acquiring the necessary domain knowledge to be able to apply the taxonomies. Hence, we could not evaluate reliability as we identified no internal measurement for this quality attribute. However, other evaluators might be able to source such data as many external measures are simple counts, e.g. unclassified objects (comprehensiveness), misclassified objects (robustness), unused constructs (conciseness), classification differences between classifiers (mutual exclusiveness and reliability). If such data is not available, the use of internal measurements might be an acceptable trade-off for comparing taxonomies \emph{before} use.

\begin{table*}[t]
    \caption{Comparison of domain specific taxonomies}\label{tab:eval}
    \begin{threeparttable}
    \centering
    \footnotesize
    \begin{tabular}{lp{4cm}llllll}
    \toprule
         Quality attribute & Measurement & \multicolumn{2}{c}{Construction} & \multicolumn{2}{c}{Cyber security} & \multicolumn{2}{c}{Economic activities}  \\ 
         & & Uniclass & Omniclass & EUCT & Mahaini & NAICS & NACE \\
    \midrule
         \multirow{4}{*}{Comprehensiveness} & Does the taxonomy originate from a diverse set of data sources? & Yes & Yes & Yes & Yes & Yes & No data \\
         & Do the taxonomy creators have a diverse background? & Yes & Yes & No & No & Yes & Yes \\
         & Were diverse data analysis methods used to create the taxonomy? & No data & No data & No & No & No data & No data \\
         & Has the taxonomy been evaluated by external experts? & Yes & Yes & Yes & No & Yes & Yes \\
         \midrule
         Robustness & $R(T)$ & 0.63 & 0.79 & 0.31 & 0.57 & 0.81 & 0.77 \\
         \midrule
         \multirow{5}{*}{Conciseness} & Dimensions & 12 & 11 & 3 & 1 & 1 & 1 \\
         & Categories & 1,827 & 4,084 & 18 & 425 & 613 & 244 \\
         & Characteristics & 12,501 & 14,569 & 188 & 1,529 & 1,056 & 615 \\
         & Maximum depth & 5 & 8 & 3 & 10 & 6 & 5 \\
         & $C(T)$ & 0.11 & 0.11 & 0.19 & 0.14 & 0.15 & 0.16 \\
         \midrule
         \multirow{5}{*}{Extensibility} & Change process covers the following constructs\tnote{a} & No data & ca, ch & No data & No data & No data & No data \\
         & Change process covers the following change types\tnote{b} & No data & a, m & No data & No data & No data & a \\
         & Mandate to change the taxonomy & Owners/ext & Owners/ext & No data & Owners/ext & Owners/ext & Owners \\
         & Customization of the taxonomy & No data & No data & No data & No data & No data & No data \\
         & $RoC_{2021-05-04} $ & 0.01079 & 0.00018 & 0.00066 & 0 & 0.00046 & 0.00016 \\
         \midrule
         \multirow{2}{*}{Explanatory} & Structuring elements besides dimensions? & No data & No data & No data & No data & No data & No data \\
         & Support for choosing dimensions or categories?\tnote{a} & di & di ca & di ca & ca & di ca & di ca \\
         \midrule
         Mutual exclusiveness & Classification constraint & No data & No data & No data & No data & yes & yes \\
         \midrule
         Reliability & \multicolumn{7}{l}{We did not evaluate this attributes since it requires the application of the taxonomy.} \\
    \bottomrule 
    \end{tabular}
    \begin{tablenotes}
        \item[a] di = dimension, ca = category, ch = characteristic
        \item[b] a = addition, m = modification, d = deletion
    \end{tablenotes}
    \end{threeparttable}
\end{table*}

\subsection{Construction}
The main purpose of taxonomies in the construction industry is to organize and exchange project information, cost estimation, and in building information modeling (BIM)~\citep{afsari_comparison_2016}. We are interested in using such taxonomies in order to support engineers tracing requirements to downstream artifacts, such as digital construction models. The goal is to enable more efficient, complete and requirements based verification, which is often not possible with thousands of requirements and tens of thousands modeled objects. In our previous work~\citep{unterkalmsteiner_early_2020}, we developed a recommendation system that suggested engineers which classes from a classification system are likely to be relevant for a particular system requirement in a infrastructure project. We are now interested to compare different classification systems and evaluate which one has advantageous characteristics that are desirable when implementing such a text- and hierarchy-based recommender.

Both Omniclass (North America) and Uniclass (United Kingdom) have seen widespread adoption in industry~\citep{afsari_comparison_2016}. Table~\ref{tab:eval} summarizes the evaluation and we discuss each quality attribute next.

\emph{Robustness.} According to our evaluation, Omniclass has a higher robustness ($R(T)=0.79$) than Uniclass ($R(T)=0.63$). This is remarkable since Omniclass has also a higher number of categories and characteristics than Uniclass. This indicates, at least if we assume that our measure is a good representation of robustness, that robustness does not have to be necessarily traded-off with conciseness. Furthermore, this indicates that the content in Omniclass is more cohesive and orthogonal than the content in Uniclass.

Our analysis also identifies the leaf node groups that exhibit the lowest robustness, i.e. are not cohesive and orthogonal (see our definition in Section~\ref{sec:robmes}). Omniclass has 2,056 leaf node groups that we analyzed. Table~\ref{tab:robdetails} shows the top and bottom five groups. The first number in the group column (21, 22, 23, 13, 49 and 36) indicates the dimension from which the leaf node group originates. Looking at the results, dimensions 21 and 22 seem to be more robust than, for example, dimension 13. Indeed, a more detailed analysis shows that the count of leaf node groups in the bottom-third is highest in dimension 13 (77\%), followed by 34 (71\%) and 49 (65\%) (see Table~\ref{tab:omniclass_rb_analysis}).

\begin{table}[t]
    \caption{Omniclass robustness analysis}
    \begin{subtable}[h]{0.45\textwidth}
    \centering
    \footnotesize
    \begin{tabular}{llll}
    \toprule
    Group & \# nodes & \# outside nodes & Outside proportion \\
    \midrule
        21-01 20 10 & 3 & 0 & 0.0 \\
        22-02 51 33 & 3 & 0 & 0.0 \\
        22-03 01 30 & 4 & 0 & 0.0 \\
        22-03 01 40 & 4 & 0 & 0.0 \\
        22-03 21 21 & 3 & 0 & 0.0 \\
        \ldots & \ldots & \ldots & \ldots \\
        23-13 23 11 13 & 8 & 93,411 & 0.978 \\
        13-63 13 & 10 & 117,023 & 0.980 \\
        49-91 11 & 31 & 362,153 & 0.980 \\
        13-51 61 & 34 & 397,121 & 0.980 \\
        36-51 73 11 13 11 & 9 & 105,349 & 0.980 \\
    \bottomrule 
    \end{tabular}
    \caption{Robustness results of top and bottom five leaf node groups in Omniclass}
    \label{tab:robdetails}
    \end{subtable}
    \hfill
    \begin{subtable}[h]{0.45\textwidth}
    \centering
    \footnotesize
    \begin{tabular}{llll}
    \toprule
    Dimension & Count & Bottom-third Count & Percentage\\
    \midrule
    13 & 78	& 60 & 77\\
    34 & 14 & 10 & 71\\
    49 & 99 & 64 & 65\\
    41 & 26 & 16 & 62\\
    21 & 89 & 46 & 52\\
    36 & 81 & 38 & 47\\
    12 & 22 & 10 & 45\\
    11 & 91 & 37 & 41\\
    33 & 30 & 9 & 30\\
    23 & 742 & 194 & 26\\
    22 & 784 & 203 & 26\\
    \bottomrule
    \end{tabular}
    \caption{Robustness analysis}\label{tab:omniclass_rb_analysis}
    \end{subtable}
\end{table}

\emph{Comprehensiveness.} Both taxonomies are very similar in terms of comprehensiveness. They origin from previous, more specialized taxonomies and draw from ISO 12006-2 "Organization of information about construction works"~\citep{afsari_comparison_2016}. While both taxonomies were developed by committees, they received input from users on draft publications~\citep{gelder_design_2015, construction_specifications_institute_contributors_2021}. Hence we can argue that the taxonomy developers have a diverse background. We did not find any information on whether diverse analysis methods were used in taxonomy construction. Finally, both taxonomies seem to foresee an external review process by experts. In Omniclass, the dimensions are individually approved as National Standards~\citep{the_construction_specifications_institute_omniclass_2019} and Uniclass is open for comments and suggestions from users~\citep{gelder_design_2015}.

\emph{Mutual exclusiveness.} We did not find any information in the taxonomy descriptions whether mutual exclusiveness is enforced.

\emph{Conciseness.} Omniclass and Uniclass exhibit the same conciseness ($C(T)=0.11$), even though they differ in number of dimensions (11 vs. 12), categories (4,084 vs 1,827) and characteristics (14,569 vs. 12,501). Since the proposed conciseness measure takes the depth of categories and characteristics into account, the higher number of characteristics in Omniclass is compensated by the broader tree, i.e. Omniclass has more categories than Uniclass.
%TODO Add here table showing conciseness of the dimensions of both taxonomies. If there is still space and if the additional data can be discussed.

\emph{Explanatory.} Omniclass provides descriptions for each dimension and category in their user guidelines~\citep{the_construction_specifications_institute_omniclass_2019}. Uniclass, on the other hand, provides only a brief explanation of dimensions~\citep{nbs_what_2021}. While the descriptions are useful, they provide little direct support for classification since both taxonomies have a high number of categories, making it rather impossible for a human user to benefit from them. On the other hand, the additional information in the descriptions can be used for text mining and analysis, which could be beneficial for applications where the taxonomies are used to create decisions support for engineers.

\emph{Extensibility.} We could not find any information about the change process or customization possibilities for Uniclass. Omniclass allows to modify and add both categories and characteristics and describe the process in their user guideline~\citep{the_construction_specifications_institute_omniclass_2019}. Changes to the taxonomies are possible by the owners which allow external input from users~\citep{the_construction_specifications_institute_omniclass_2019, nbs_uniclass_2021}. Finally, the rate of change indicates that modifications are done more frequently in UniClass.

\paragraph{Decision}
Based on the robustness score, Omniclass should be chosen over Uniclass. While there is little difference in comprehensiveness, mutual exclusiveness and conciseness between the two, Omniclass is also preferable w.r.t. explanatory power as it provides also category descriptions in addition to dimension descriptions. Looking at particular dimensions in Omniclass, we observe that there is a varying degree of robustness. It would likely be beneficial (in terms of recommender accuracy) to exclude dimensions with the highest percentage of low robustness leaf node groups (see Table~\ref{tab:omniclass_rb_analysis}), if they are not relevant for the particular use case. 

\subsection{Cyber security}
We compare two diverse taxonomies for cyber security. The European Cybersecurity Taxonomy (hereafter referred to as EUCT) has the goal of aligning cyber security terminologies and definitions to facilitate the categorization of cyber security competences~\citep{nai_fovino_proposal_2019}. The taxonomy was developed following the five steps of (1) defining the subject scope, (2) identifying data sources, (3) collecting terms and concepts, (4) clustering concepts, and finally (5) conciliating content and structure of the taxonomy. This resulted in a taxonomy with three dimensions, 18 categories and 188 characteristics.

\citet{mahaini_building_2019} present and illustrate a different taxonomy development approach (hereafter referred to as Mahaini). They suggest a co-development process in which human experts are supported by natural language processing and information retrieval tools, generating a taxonomy from relevant textual documents.

Both EUCT and Mahaini have defined use cases for their taxonomies. EUCT is used to classify cyber security competencies of organizations in the EU while Mahaini suggests that the taxonomy can be used to classify cyber security related discussions in online social networks or other textual sources. Table~\ref{tab:eval} summarizes the evaluation of both taxonomies.

\begin{table}[t]
    \caption{Examples with low robustness in EUCT}
    \label{tab:cohesivenessEUCT}
    \centering
    \footnotesize
    \begin{tabular}{p{1.5cm}p{10cm}ll}
    \toprule
    Dimension & Characteristics & \# outside nodes & Outside proportion \\
    \midrule
    Sectors & Audiovisual and media, Chemical, Defence, Digital Services and Platforms, Energy, Financial, Food and drink, Government, Health, Manufacturing and Supply Chain, Nuclear, Safety and Security, [\ldots] & 2517 & 0.970\\
    Technologies and Use Cases & Artificial intelligence, Big Data, Blockchain and Distributed Ledger Technology (DLT), Cloud, Edge and Virtualisation, Critical Infrastructure Protection (CIP), Protection of public spaces, Disaster resilience and crisis management, [\ldots] & 3752 & 0.989\\
    \bottomrule 
    \end{tabular}
\end{table}

\emph{Robustness.} Mahaini ($R(T)=0.57$) has a higher robustness than EUCT ($R(T)=0.31$). In this example, the intuition by \citet{nickerson_method_2013} that a taxonomy with more categories and characteristics is also more robust, correlates with our robustness measurement.
Looking at the leaf node groups in EUCT that exhibit low robustness, we find that two dimensions are only two levels deep, and are, by their nature, not very cohesive. Table~\ref{tab:cohesivenessEUCT} shows the characteristics of these two dimensions, which are, while related through the grouping of the dimension, quite diverse in their content. One could also argue that technologies and use cases should be in two separate dimensions.

\emph{Comprehensiveness.} Looking at the measurements we defined, EUCT seems to be more comprehensive than Mahaini. Both taxonomies draw from various data sources for their development. We looked up the profiles of the authors that published the taxonomies. While Mahaini was developed by three authors, the two first authors worked at the same university. EUCT was developed by seven authors, combining experience from different universities and activities in the European commission. In both cases, no expertise from industry or from other organizations specialized in cyber security was directly involved in the development of the taxonomy. While Mahaini propose a new taxonomy development method, it was, as also in EUCT, the only method they applied. Finally, EUCT's first draft has been reviewed by project partners including companies and universities. Such a review has not been reported for Mahaini.

\emph{Mutual exclusiveness.} We did not find any information in the taxonomy descriptions whether mutual exclusiveness is enforced.

\emph{Conciseness.} EUCT ($C(T)=0.19$) has a higher conciseness than Mahaini ($C(T)=0.14$). Mahaini's higher number of characteristics (1,529 vs. 188) and depth (10 vs. 3) might be attributed to the fact that it was developed with the purpose of using it for automated classification, while EUCT is applied by human classifiers.

\emph{Explanatory.} EUCT provides descriptions for each dimension and category~\citep{nai_fovino_proposal_2019}. Mahaini provides descriptions of the top-level categories of the taxonomy~\citep{mahaini_building_2019}. There are no further aids for the explanatory power of the taxonomies.

\emph{Extensibility.} While neither EUCT nor Mahaini define a change process, EUCT acknowledges that defining a taxonomy in this domain is a moving target and needs to be open for modifications~\citep{nai_fovino_proposal_2019}. Mahaini was developed as an illustration for a particular taxonomy development approach. The authors welcome suggestions from the community, do however not focus on how the taxonomy can be modified in the future~\citep{mahaini_building_2019}. 

\paragraph{Decision}
Based on the robustness score, Mahaini should be chosen over EUCT, even if comprehensiveness is favorable in EUCT. The robustness analysis of EUCT pinpoints a potential critical issue with the classification system: the dimensions "Sectors" and "Technologies and Use Cases" contain concepts that are not cohesive (see Table~\ref{tab:cohesivenessEUCT}). If EUCT must be used for classification, one might consider excluding those dimensions.

\subsection{Economic activities}
The North American Industry Classification System (hereafter referred to as NAICS) and the European Classification of Economic Activities (hereafter referred to as NACE\footnote{The acronym originates from the French name of the taxonomy: \emph{N}omenclature générale des \emph{A}ctivités économiques dans les \emph{C}ommunautés \emph{E}uropéennes.}) are a fundamental tool to produce statistics on economic activities on the national, European and world-wide level. Both taxonomies aim to provide a comprehensive system with mutually exclusive categories~\citep{european_commission_nace_2008,executive_office_of_the_president_office_of_management_and_budget_north_2017}.

Use cases of these taxonomies for software engineering practitioners are likely related to the automated classification of data, for example, of streams of information that were originally not intended to be used for statistical analysis. Table~\ref{tab:eval} summarizes the evaluation of both taxonomies.

\begin{table}[t]
    \caption{Examples of high and low robustness in NAICS}
    \label{tab:cohesivenessNAICS}
    \centering
    \footnotesize
    \begin{tabular}{p{3cm}p{8.5cm}ll}
    \toprule
    Category & Characteristics & \# outside nodes & Outside proportion \\
    \midrule
    Electric Power Generation & Hydroelectric Power Generation, Fossil Fuel Electric Power, Generation, Nuclear Electric Power Generation, Solar Electric Power Generation, Wind Electric Power Generation, [\ldots] & 0 & 0\\
    \midrule
    Nonferrous Metal Foundries & Nonferrous Metal Die-Casting Foundries, Aluminum Foundries (except Die-Casting), Other Nonferrous Metal Foundries (except Die-Casting) & 0 & 0 \\
    \midrule
    Other Insurance Related Activities & Claims Adjusting, Third Party Administration of Insurance and Pension Funds, All Other Insurance Related Activities & 881 & 0.454 \\
    \midrule
    Other Personal Services & Pet Care (except Veterinary) Services, Photofinishing, Parking Lots and Garages, Parking Lots and Garages, [\ldots] & 1335 & 0.688 \\
    \bottomrule 
    \end{tabular}
\end{table}
\emph{Robustness.} We observe a similar robustness between NAICS ($R(T)=0.81$) and NACE ($R(T)=0.77$) that is higher than in the previously studied taxonomies (except for Omniclass). Table~\ref{tab:cohesivenessNAICS} exemplifies four leaf group nodes from NAICS that exhibit high (no outside nodes) and low robustness (many outside nodes). Comparing the categories, one can observe that semantically related characteristics form cohesive groups, while "catch-all" categories ("Other...") seem to contain less cohesive characteristics.

\emph{Comprehensiveness.} Both taxonomies are very similar in terms of comprehensiveness, and are, based on our criteria, to a large degree comprehensive. We could not find information about the data sources that were used to create NACE in the documentation of the taxonomy~\citep{european_commission_nace_2008}. NAICS was developed with input from Mexico, Canada and the United States. Both taxonomies were reviewed by external committees.

\emph{Mutual exclusiveness.} Since businesses may engage in several activities, both taxonomies describe how to determine the main economic activities of a business in order to achieve a unique classification. 

\emph{Conciseness.} NACE ($C(T)=0.16$) and NAICS ($C(T)=0.15$) have a comparable conciseness whereby the simpler taxonomy seems also to exhibit lower robustness.  

\emph{Explanatory.} Both taxonomies are accompanied with rich descriptions of dimensions and categories. NAICS provides a brief description on how to determine an businesses' industry classification~\citep{executive_office_of_the_president_office_of_management_and_budget_north_2017} whereas NACE dedicates a whole chapter on classification rules~\citep{european_commission_nace_2008}.

\emph{Extensibility.} We were not able to find information about the change process in both taxonomies. NAICS is reviewed every five years to reflect changes in the economy and change proposals from external interested parties are accepted ~\citep{executive_office_of_the_president_office_of_management_and_budget_north_2017}. NACE has left gaps in the numerical coding system where they foresaw likely additions in the future and can be customized for particular national requirements~\citep{european_commission_nace_2008} (even though we did not find information which changes are possible). 
We observe a higher $RoC$ with NAICS, which has seen five releases since 1997 (every five years), while NACE saw four releases since 1970.

\paragraph{Decision}
Based on the robustness score, NAICS should be chosen over NACE, even though the difference between the two taxonomies is less pronounced than in the previous examples. Looking at the categories with a high number of outside nodes (see Table~\ref{tab:cohesivenessNAICS}), one might consider to exclude "catch-all" categories that contain "other" concepts from recommender training.

\section{Discussion}\label{sec:discussion}
In this section, we discuss the main observations we made evaluating the six taxonomies, the lessons learned from applying the measurements, and their limitations.

To understand the objective internal measurement results of robustness better, we analyzed those parts of the taxonomies that were scored with low robustness. For example, looking at Omniclass we could identify dimensions that exhibited low robustness and could therefore be discarded form the inclusion in a recommender system. Another example are the incohesive dimensions in EUCT. The low robustness of EUCT, compared to Mahaini, was surprising given the dedicated effort put into the taxonomy, while Mahaini is a "side-product" resulting from the proposal of a novel taxonomy development approach. We made a similar observation in NAICS categories that exhibited low robustness, named "Other...", that contained rather unrelated characteristics. Hence, we conclude that the objective internal robustness measurement can provide support for locating elements in (large) taxonomies that may result in lower classification performance due to incohesiveness of its elements, both for human and computational classifiers.

We also observe that, generally, robustness is traded off for conciseness. The taxonomies in the construction domain were a surprising exception.
Taxonomies created by committees or large interest groups (construction and economic activities) tend to be less concise than taxonomies created by smaller teams (cyber security). However, within the cyber security domain, the larger team created the more concise EUCT. This can be explained by the semi-automated method Mahaini was constructed, that led to a much more fine-grained and deeper classification tree (see conciseness measurements in Table~\ref{tab:eval}).  

Overall, we found that by prioritizing the quality attributes based on the (hypothetical) use case of implementing a recommendation system, we could make an informed decision on selecting a taxonomy, based on objective criteria. Hence, we found the measurements useful and efficient, in particular those ones that can be calculated automatically and do not rely on manual data extraction.

\subsection{Lessons learned}
It was seldom the case that we could find the data for subjective internal measurements directly in the documentation of the taxonomies under a clear header. For example, to evaluate \emph{comprehensivness}, one of the questions is \emph{Do the taxonomy creators have a diverse background?} To answer this question, we had to find the authors of the taxonomy and assess their background (EUCT), or interpret that a list of over 100 companies contributed to the development of a taxonomy represents a group with diverse background (Omniclass). 

% lack of documentation
The extraction of subjective internal measurements depends heavily on the supporting information that accompanies the taxonomy. This information is usually found to a larger extent in taxonomies with resources available for documentation, that are in active use (Omniclass, Uniclass) or developed by/for governments (NACE, NAICS, EUCT), compared to taxonomies created by research groups (Mahaini). Still, evaluating the subjective internal measures is not resource-intensive (if the information can be found). The extraction by a non-expert in the field took a median of 3 hours per taxonomy. This time varies depending on how well the taxonomy is documented.

The implementation of the objective internal measurements requires that the taxonomies are available in a machine-readable format. All taxonomies were published by their owners in such formats (csv, xls, json), except for EUCT which we had to transcribe manually from the published report (pdf), and two dimensions from Omniclass which we excluded from the evaluation.
The remaining dimensions in Omniclass contained also several typos in the coding scheme, leading us to assume that the tables are maintained manually. We used the coding scheme to generate the classification tree and had to correct these inconsistencies manually. In NAICS, three categories with the same name were grouped together under one code (31-33 Manufacturing, 45-45 Retail Trade, 48-49 Transportation and Warehousing), which is the only exception in the otherwise perfect tree structure of the coding scheme. This makes us suspect that machine-readability of the raw data was not a priority for the taxonomy's creators.

For each taxonomy, we implemented a parser that extracted the information from the source artifacts and created a uniform tree representation on which the measurements can be applied. The source code is available in the supplementary material~\citep{unterkalmsteiner_supp_2021}.

\subsection{Limitations}
In the validity threats to this study (Section~\ref{sec:vt}) we discussed how to improve the generalizability of our results. While we are confident that our strategy helped to cover a majority of the relevant studies on taxonomy quality from the information systems and software engineering fields, it did not mitigate the risk of missing relevant papers \emph{outside} of these fields. For example, the semantic web community has researched extensively the evaluation of ontologies~\citep{gangemi2006modelling, obrst2007evaluation, hlomani2014approaches, raad2015survey, mcdaniel2019evaluating}. None of these works have been cited in the studies that evaluated taxonomy quality and that we reviewed in our study. This is to some extent explainable by the more powerful mechanisms for knowledge representation of ontologies compared to taxonomies~\citep{gruninger_ontology_2008}. Taxonomies, in essence, aim to represent hierarchical relationships, while ontologies are able to model more complex relationships~\citep{garshol_metadata_2004}. Consequently, the evaluation approaches for ontologies differ from those for taxonomies, even though we observe overlaps. For example, \citet{gangemi2006modelling} describe three dimensions of ontology evaluation measures: (1) measures of structural properties of ontologies; some of these are also applicable to taxonomies, e.g. depth of a node, while others make only sense in an ontology due to its graph representation, e.g. tangledness of a node. (2) Measures of functional properties of ontologies; again, some seem to correspond to quality attributes in taxonomies, e.g. the proportion of agreement that experts have on the elements of an ontology is similar to the quality attribute of reliability which we described in Section~\ref{sec:reliability}. (3) Measures of the usability profile of ontologies that define measurements for the practical use of ontologies; these measurements describe metadata about an ontology and its elements, such as the ease of access to ontology use instructions. 
To summarize, several but not all aspects of ontology evaluation could be applied also to the evaluation of taxonomy quality in information systems and software engineering. There is a clear need and potential of pursuing research that bridges these fields. We proceed now with the discussion of limitations of particular measurements proposed in our study.

The extraction of subjective internal measurements depends on the availability of information in the documentation accompanying the taxonomy. In our summary in Table~\ref{tab:eval}, we marked measurements with "no data" when we could not find the information required to answer our questions. This does not mean that the aspect in question does not exist at all, but rather we could not find pertinent information. The answers for comprehensiveness questions seem to be the easiest to extract while extensibility seems to be less documented. Since we did not have any data on the application of the different taxonomies, we could not evaluate reliability (Section~\ref{sec:reliability}) nor evaluate any other external measurement (see Table~\ref{tab:qam}).

In order to implement the robustness measure, we created a doc2vec model from the Wikipedia article corpus. We suspect that a domain specific corpus for each taxonomy would lead to more accurate results~\citep{nooralahzadeh_evaluation_2018}. Furthermore, our robustness measure does not take into consideration faceted taxonomies (e.g. as in Uniclass and Omniclass). In a faceted taxonomy is it possible to assign an object to multiple dimensions without violating the mutual exclusiveness constraint. Hence, the robustness measurement should consider only categories and characteristics \emph{within} a dimension.

% EUCT contains two dimensions that contain rather incohesive categories, based on the measurement we propose. The sector dimensions contains various sectors that are not related, hence the low cohesiveness. The technologies and use cases dimension is however indeed a mixture of two concepts and should be refined.

We do not suggest any thresholds for the objective internal measurements as we did not correlate those measurements with, for example, the perceived usefulness of a taxonomy or show a cause-effect relationship to external quality attributes. The definitions of such thresholds could be done in future work. However, we want to emphasize that the main benefit of performing an evaluation with the suggested measurements is the \emph{relative} comparison of either different taxonomies, or iterations of the same taxonomy after it has seen improvements over time. For such evaluations, thresholds are not necessary.

\section{Conclusions and future work}\label{sec:cfw}
Taxonomies are a fundamental tool to synthesize domain specific knowledge that can be used by humans for classification as well as by computers to support engineers in solving knowledge intensive problems. Creating high quality taxonomies is therefore of central importance. \citet{nickerson_method_2013} proposed five quality attributes for taxonomies (conciseness, robustness, comprehensiveness, extensibility and explanatory), did however leave concrete measurements how these attributes can be evaluated to future work. In this paper, we add two more quality attributes (mutual exclusiveness, reliability) to Nickerson et al.'s set and suggest concrete measurements for each of the seven quality attributes. 
We found that evaluating the six taxonomies led to useful insights that can be used to base decisions on which taxonomy to use for a particular purpose. To illustrate this, we chose the purpose of creating a recommendation system, prioritizing robustness, comprehensiveness and mutual exclusiveness over the other attributes that would likely be more important if the taxonomies were applied by a human user. 

While the quality attributes originate from the software engineering and information systems literature, they are not domain specific and we expect them, based on our experience applying them, to be useful in a wide variety of scenarios. However, we see also the need to adapt, integrate and use ontology evaluation approaches that originate from the semantic web community.   

The causal relationship between the proposed internal measurements and the external quality attributes still need to be established empirically. As a first step, existing data (unclassified/misclassified objects) could be used to investigate correlations between internal measurements, such as semantic proximity and distance of taxonomy constructs, and external quality attributes, such as robustness. Then, in a second step, controlled experiments could be designed to investigate the causal relationship between internal measurements and external attributes. Such studies would be very valuable as external measurements are expensive to collect and therefore often impractical for the goal of choosing between taxonomies in a particular domain.

\section{Acknowledgments}
The work in this paper was partly funded by projects D-CAT (Trafikverket) and S.E.R.T. (KK-stiftelsen).

\paragraph{Conflict of Interest}
The authors declare that they have no conflict of interest.

\bibliography{references}

\begin{thebibliography}{}

\bibitem [\protect \citeauthoryear {%
Afsari%
\ \BBA {} Eastman%
}{%
Afsari%
\ \BBA {} Eastman%
}{%
{\protect \APACyear {2016}}%
}]{%
afsari_comparison_2016}
\APACinsertmetastar {%
afsari_comparison_2016}%
\begin{APACrefauthors}%
Afsari, K.%
\BCBT {}\ \BBA {} Eastman, C\BPBI M.%
\end{APACrefauthors}%
\unskip\
\newblock
\APACrefYearMonthDay{2016}{}{}.
\newblock
{\BBOQ}\APACrefatitle {A comparison of construction classification systems used
  for classifying building product models} {A comparison of construction
  classification systems used for classifying building product models}.{\BBCQ}
\newblock
\BIn{} \APACrefbtitle {52nd {ASC} {Annual} {International} {Conference}
  {Proceedings}} {52nd {ASC} {Annual} {International} {Conference}
  {Proceedings}}\ (\BPGS\ 1--8).
\PrintBackRefs{\CurrentBib}

\bibitem [\protect \citeauthoryear {%
Aksulu%
\ \BBA {} Wade%
}{%
Aksulu%
\ \BBA {} Wade%
}{%
{\protect \APACyear {2010}}%
}]{%
aksulu_comprehensive_2010}
\APACinsertmetastar {%
aksulu_comprehensive_2010}%
\begin{APACrefauthors}%
Aksulu, A.%
\BCBT {}\ \BBA {} Wade, M.%
\end{APACrefauthors}%
\unskip\
\newblock
\APACrefYearMonthDay{2010}{{\APACmonth{11}}}{}.
\newblock
{\BBOQ}\APACrefatitle {A {Comprehensive} {Review} and {Synthesis} of {Open}
  {Source} {Research}} {A {Comprehensive} {Review} and {Synthesis} of {Open}
  {Source} {Research}}.{\BBCQ}
\newblock
\APACjournalVolNumPages{Journal of the Association for Information
  Systems}{11}{11}{576--656}.
\PrintBackRefs{\CurrentBib}

\bibitem [\protect \citeauthoryear {%
Almufareh%
, Abaoud%
\BCBL {}\ \BBA {} Moniruzzaman%
}{%
Almufareh%
\ \protect \BOthers {.}}{%
{\protect \APACyear {2018}}%
}]{%
almufareh_taxonomy_2018}
\APACinsertmetastar {%
almufareh_taxonomy_2018}%
\begin{APACrefauthors}%
Almufareh, M.%
, Abaoud, D.%
\BCBL {}\ \BBA {} Moniruzzaman, M.%
\end{APACrefauthors}%
\unskip\
\newblock
\APACrefYearMonthDay{2018}{{\APACmonth{06}}}{}.
\newblock
{\BBOQ}\APACrefatitle {Taxonomy {Development} for {Virtual} {Reality} ({VR})
  {Technologies} in {Healthcare} {Sector}} {Taxonomy {Development} for
  {Virtual} {Reality} ({VR}) {Technologies} in {Healthcare} {Sector}}.{\BBCQ}
\newblock
\BIn{} \APACrefbtitle {International {Conference} on {Design} {Science}
  {Research} in {Information} {Systems} and {Technology}} {International
  {Conference} on {Design} {Science} {Research} in {Information} {Systems} and
  {Technology}}\ (\BPGS\ 146--156).
\newblock
\APACaddressPublisher{Chennai, India}{Springer International Publishing}.
\PrintBackRefs{\CurrentBib}

\bibitem [\protect \citeauthoryear {%
Alrige%
\ \BBA {} Chatterjee%
}{%
Alrige%
\ \BBA {} Chatterjee%
}{%
{\protect \APACyear {2015}}%
}]{%
alrige_toward_2015}
\APACinsertmetastar {%
alrige_toward_2015}%
\begin{APACrefauthors}%
Alrige, M.%
\BCBT {}\ \BBA {} Chatterjee, S.%
\end{APACrefauthors}%
\unskip\
\newblock
\APACrefYearMonthDay{2015}{{\APACmonth{05}}}{}.
\newblock
{\BBOQ}\APACrefatitle {Toward a {Taxonomy} of {Wearable} {Technologies} in
  {Healthcare}} {Toward a {Taxonomy} of {Wearable} {Technologies} in
  {Healthcare}}.{\BBCQ}
\newblock
\BIn{} \APACrefbtitle {International {Conference} on {Design} {Science}
  {Research} in {Information} {Systems} and {Technology}} {International
  {Conference} on {Design} {Science} {Research} in {Information} {Systems} and
  {Technology}}\ (\BPGS\ 496--504).
\newblock
\APACaddressPublisher{Dublin, Ireland}{Springer International Publishing}.
\PrintBackRefs{\CurrentBib}

\bibitem [\protect \citeauthoryear {%
Bapna%
, Goes%
, Gupta%
\BCBL {}\ \BBA {} Jin%
}{%
Bapna%
\ \protect \BOthers {.}}{%
{\protect \APACyear {2004}}%
}]{%
bapna_user_2004}
\APACinsertmetastar {%
bapna_user_2004}%
\begin{APACrefauthors}%
Bapna, R.%
, Goes, P.%
, Gupta, A.%
\BCBL {}\ \BBA {} Jin, Y.%
\end{APACrefauthors}%
\unskip\
\newblock
\APACrefYearMonthDay{2004}{}{}.
\newblock
{\BBOQ}\APACrefatitle {User {Heterogeneity} and {Its} {Impact} on {Electronic}
  {Auction} {Market} {Design}: {An} {Empirical} {Exploration}} {User
  {Heterogeneity} and {Its} {Impact} on {Electronic} {Auction} {Market}
  {Design}: {An} {Empirical} {Exploration}}.{\BBCQ}
\newblock
\APACjournalVolNumPages{MIS Quarterly}{28}{1}{21--43}.
\PrintBackRefs{\CurrentBib}

\bibitem [\protect \citeauthoryear {%
Basciani%
, Di~Rocco%
, Di~Ruscio%
, Iovino%
\BCBL {}\ \BBA {} Pierantonio%
}{%
Basciani%
\ \protect \BOthers {.}}{%
{\protect \APACyear {2019}}%
}]{%
basciani_tool-supported_2019}
\APACinsertmetastar {%
basciani_tool-supported_2019}%
\begin{APACrefauthors}%
Basciani, F.%
, Di~Rocco, J.%
, Di~Ruscio, D.%
, Iovino, L.%
\BCBL {}\ \BBA {} Pierantonio, A.%
\end{APACrefauthors}%
\unskip\
\newblock
\APACrefYearMonthDay{2019}{}{}.
\newblock
{\BBOQ}\APACrefatitle {A tool-supported approach for assessing the quality of
  modeling artifacts} {A tool-supported approach for assessing the quality of
  modeling artifacts}.{\BBCQ}
\newblock
\APACjournalVolNumPages{}{51}{}{173--192}.
\PrintBackRefs{\CurrentBib}

\bibitem [\protect \citeauthoryear {%
Basciani%
, Rocco%
, Ruscio%
, Iovino%
\BCBL {}\ \BBA {} Pierantonio%
}{%
Basciani%
\ \protect \BOthers {.}}{%
{\protect \APACyear {2016}}%
}]{%
basciani_customizable_2016}
\APACinsertmetastar {%
basciani_customizable_2016}%
\begin{APACrefauthors}%
Basciani, F.%
, Rocco, J\BPBI d.%
, Ruscio, D\BPBI d.%
, Iovino, L.%
\BCBL {}\ \BBA {} Pierantonio, A.%
\end{APACrefauthors}%
\unskip\
\newblock
\APACrefYearMonthDay{2016}{}{}.
\newblock
{\BBOQ}\APACrefatitle {A Customizable Approach for the Automated Quality
  Assessment of Modelling Artifacts} {A customizable approach for the automated
  quality assessment of modelling artifacts}.{\BBCQ}
\newblock
\BIn{} \APACrefbtitle {10th International Conference on the Quality of
  Information and Communications Technology ({QUATIC})} {10th international
  conference on the quality of information and communications technology
  ({QUATIC})}\ (\BPGS\ 88--93).
\PrintBackRefs{\CurrentBib}

\bibitem [\protect \citeauthoryear {%
Basili%
, Caldiera%
\BCBL {}\ \BBA {} Rombach%
}{%
Basili%
\ \protect \BOthers {.}}{%
{\protect \APACyear {1994}}%
}]{%
basili_goal_1994}
\APACinsertmetastar {%
basili_goal_1994}%
\begin{APACrefauthors}%
Basili, V\BPBI R.%
, Caldiera, G.%
\BCBL {}\ \BBA {} Rombach, H\BPBI D.%
\end{APACrefauthors}%
\unskip\
\newblock
\APACrefYearMonthDay{1994}{}{}.
\newblock
{\BBOQ}\APACrefatitle {The goal question metric approach} {The goal question
  metric approach}.{\BBCQ}
\newblock
\APACjournalVolNumPages{Encyclopedia of Software Engineering}{1}{}{528--532}.
\PrintBackRefs{\CurrentBib}

\bibitem [\protect \citeauthoryear {%
Beevi%
, Wagner%
, Hallerstede%
\BCBL {}\ \BBA {} Pedersen%
}{%
Beevi%
\ \protect \BOthers {.}}{%
{\protect \APACyear {2015}}%
}]{%
beevi_data_2015}
\APACinsertmetastar {%
beevi_data_2015}%
\begin{APACrefauthors}%
Beevi, F\BPBI H\BPBI A.%
, Wagner, S.%
, Hallerstede, S.%
\BCBL {}\ \BBA {} Pedersen, C\BPBI F.%
\end{APACrefauthors}%
\unskip\
\newblock
\APACrefYearMonthDay{2015}{{\APACmonth{11}}}{}.
\newblock
{\BBOQ}\APACrefatitle {Data quality oriented taxonomy of ambient assisted
  living systems} {Data quality oriented taxonomy of ambient assisted living
  systems}.{\BBCQ}
\newblock
\BIn{} \APACrefbtitle {{IET} {International} {Conference} on {Technologies} for
  {Active} and {Assisted} {Living} ({TechAAL}).} {{IET} {International}
  {Conference} on {Technologies} for {Active} and {Assisted} {Living}
  ({TechAAL}).}
\newblock
\APACaddressPublisher{London, UK}{IET}.
\PrintBackRefs{\CurrentBib}

\bibitem [\protect \citeauthoryear {%
Bock%
\ \BBA {} Wiener%
}{%
Bock%
\ \BBA {} Wiener%
}{%
{\protect \APACyear {2017}}%
}]{%
bock_towards_2017}
\APACinsertmetastar {%
bock_towards_2017}%
\begin{APACrefauthors}%
Bock, M.%
\BCBT {}\ \BBA {} Wiener, M.%
\end{APACrefauthors}%
\unskip\
\newblock
\APACrefYearMonthDay{2017}{{\APACmonth{12}}}{}.
\newblock
{\BBOQ}\APACrefatitle {Towards a {Taxonomy} of {Digital} {Business} {Models}
  – {Conceptual} {Dimensions} and {Empirical} {Illustrations}} {Towards a
  {Taxonomy} of {Digital} {Business} {Models} – {Conceptual} {Dimensions} and
  {Empirical} {Illustrations}}.{\BBCQ}
\newblock
\BIn{} \APACrefbtitle {Proceedings of the {International} {Conference} on
  {Informtion} {Systems}.} {Proceedings of the {International} {Conference} on
  {Informtion} {Systems}.}
\newblock
\APACaddressPublisher{Seoul, South Korea}{AIS Electronic Library}.
\PrintBackRefs{\CurrentBib}

\bibitem [\protect \citeauthoryear {%
Botha%
, Weiss%
\BCBL {}\ \BBA {} Herselman%
}{%
Botha%
\ \protect \BOthers {.}}{%
{\protect \APACyear {2018}}%
}]{%
botha_towards_2018}
\APACinsertmetastar {%
botha_towards_2018}%
\begin{APACrefauthors}%
Botha, A.%
, Weiss, M.%
\BCBL {}\ \BBA {} Herselman, M.%
\end{APACrefauthors}%
\unskip\
\newblock
\APACrefYearMonthDay{2018}{{\APACmonth{08}}}{}.
\newblock
{\BBOQ}\APACrefatitle {Towards a {Taxonomy} of {mHealth}} {Towards a {Taxonomy}
  of {mHealth}}.{\BBCQ}
\newblock
\BIn{} \APACrefbtitle {International {Conference} on {Advances} in {Big}
  {Data}, {Computing} and {Data} {Communication} {Systems}} {International
  {Conference} on {Advances} in {Big} {Data}, {Computing} and {Data}
  {Communication} {Systems}}\ (\BPGS\ 1--9).
\newblock
\APACaddressPublisher{Durban, South Africa}{IEEE}.
\PrintBackRefs{\CurrentBib}

\bibitem [\protect \citeauthoryear {%
Bärenfänger%
\ \protect \BOthers {.}}{%
Bärenfänger%
\ \protect \BOthers {.}}{%
{\protect \APACyear {2016}}%
}]{%
barenfanger_classifying_2016}
\APACinsertmetastar {%
barenfanger_classifying_2016}%
\begin{APACrefauthors}%
Bärenfänger, R.%
, Drayer, E.%
, Daniluk, D.%
, Otto, B.%
, Vanet, E.%
, Caire, R.%
\BDBL {}Lisanti, B.%
\end{APACrefauthors}%
\unskip\
\newblock
\APACrefYearMonthDay{2016}{{\APACmonth{06}}}{}.
\newblock
{\BBOQ}\APACrefatitle {Classifying flexibility types in smart electric
  distribution grids: {A} taxonomy} {Classifying flexibility types in smart
  electric distribution grids: {A} taxonomy}.{\BBCQ}
\newblock
\BIn{} \APACrefbtitle {{CIRED} {Workshop}} {{CIRED} {Workshop}}\ (\BPGS\ 1--4).
\PrintBackRefs{\CurrentBib}

\bibitem [\protect \citeauthoryear {%
Chasin%
, von Hoffen%
, Cramer%
\BCBL {}\ \BBA {} Matzner%
}{%
Chasin%
\ \protect \BOthers {.}}{%
{\protect \APACyear {2018}}%
}]{%
chasin_peer--peer_2018}
\APACinsertmetastar {%
chasin_peer--peer_2018}%
\begin{APACrefauthors}%
Chasin, F.%
, von Hoffen, M.%
, Cramer, M.%
\BCBL {}\ \BBA {} Matzner, M.%
\end{APACrefauthors}%
\unskip\
\newblock
\APACrefYearMonthDay{2018}{{\APACmonth{05}}}{}.
\newblock
{\BBOQ}\APACrefatitle {Peer-to-peer sharing and collaborative consumption
  platforms: a taxonomy and a reproducible analysis} {Peer-to-peer sharing and
  collaborative consumption platforms: a taxonomy and a reproducible
  analysis}.{\BBCQ}
\newblock
\APACjournalVolNumPages{Information Systems and e-Business
  Management}{16}{2}{293--325}.
\PrintBackRefs{\CurrentBib}

\bibitem [\protect \citeauthoryear {%
Cledou%
, Estevez%
\BCBL {}\ \BBA {} Soares~Barbosa%
}{%
Cledou%
\ \protect \BOthers {.}}{%
{\protect \APACyear {2018}}%
}]{%
cledou_taxonomy_2018}
\APACinsertmetastar {%
cledou_taxonomy_2018}%
\begin{APACrefauthors}%
Cledou, G.%
, Estevez, E.%
\BCBL {}\ \BBA {} Soares~Barbosa, L.%
\end{APACrefauthors}%
\unskip\
\newblock
\APACrefYearMonthDay{2018}{{\APACmonth{01}}}{}.
\newblock
{\BBOQ}\APACrefatitle {A taxonomy for planning and designing smart mobility
  services} {A taxonomy for planning and designing smart mobility
  services}.{\BBCQ}
\newblock
\APACjournalVolNumPages{Government Information Quarterly}{35}{1}{61--76}.
\PrintBackRefs{\CurrentBib}

\bibitem [\protect \citeauthoryear {%
Conboy%
}{%
Conboy%
}{%
{\protect \APACyear {2009}}%
}]{%
conboy_agility_2009}
\APACinsertmetastar {%
conboy_agility_2009}%
\begin{APACrefauthors}%
Conboy, K.%
\end{APACrefauthors}%
\unskip\
\newblock
\APACrefYearMonthDay{2009}{{\APACmonth{08}}}{}.
\newblock
{\BBOQ}\APACrefatitle {Agility from {First} {Principles}: {Reconstructing} the
  {Concept} of {Agility} in {Information} {Systems} {Development}} {Agility
  from {First} {Principles}: {Reconstructing} the {Concept} of {Agility} in
  {Information} {Systems} {Development}}.{\BBCQ}
\newblock
\APACjournalVolNumPages{Information Systems Research}{20}{3}{329--354}.
\PrintBackRefs{\CurrentBib}

\bibitem [\protect \citeauthoryear {%
{Construction Specifications Institute}%
}{%
{Construction Specifications Institute}%
}{%
{\protect \APACyear {2021}}%
}]{%
construction_specifications_institute_contributors_2021}
\APACinsertmetastar {%
construction_specifications_institute_contributors_2021}%
\begin{APACrefauthors}%
{Construction Specifications Institute}.%
\end{APACrefauthors}%
\unskip\
\newblock
\APACrefYearMonthDay{2021}{}{}.
\newblock
\APACrefbtitle {Contributors.} {Contributors.}
\newblock
\begin{APACrefURL}
  [{2021-05-12}]\url{https://www.csiresources.org/standards/omniclass/standards-omniclass-contributors}
  \end{APACrefURL}
\PrintBackRefs{\CurrentBib}

\bibitem [\protect \citeauthoryear {%
Cugola%
\ \BBA {} Ghezzi%
}{%
Cugola%
\ \BBA {} Ghezzi%
}{%
{\protect \APACyear {1998}}%
}]{%
cugola1998software}
\APACinsertmetastar {%
cugola1998software}%
\begin{APACrefauthors}%
Cugola, G.%
\BCBT {}\ \BBA {} Ghezzi, C.%
\end{APACrefauthors}%
\unskip\
\newblock
\APACrefYearMonthDay{1998}{}{}.
\newblock
{\BBOQ}\APACrefatitle {Software Processes: a Retrospective and a Path to the
  Future} {Software processes: a retrospective and a path to the
  future}.{\BBCQ}
\newblock
\APACjournalVolNumPages{Software Process: Improvement and
  Practice}{4}{3}{101--123}.
\PrintBackRefs{\CurrentBib}

\bibitem [\protect \citeauthoryear {%
Debreceny%
, Felden%
, Ochocki%
, Piechocki%
\BCBL {}\ \BBA {} Piechocki%
}{%
Debreceny%
\ \protect \BOthers {.}}{%
{\protect \APACyear {2009}}%
}]{%
debreceny_xbrl_2009}
\APACinsertmetastar {%
debreceny_xbrl_2009}%
\begin{APACrefauthors}%
Debreceny, R.%
, Felden, C.%
, Ochocki, B.%
, Piechocki, M.%
\BCBL {}\ \BBA {} Piechocki, M.%
\end{APACrefauthors}%
\unskip\
\newblock
\APACrefYear{2009}.
\newblock
\APACrefbtitle {{XBRL} for {Interactive} {Data}: {Engineering} the
  {Information} {Value} {Chain}} {{XBRL} for {Interactive} {Data}:
  {Engineering} the {Information} {Value} {Chain}}\ (\PrintOrdinal{2009th
  edition}\ \BEd).
\newblock
\APACaddressPublisher{London ; New York}{Springer}.
\PrintBackRefs{\CurrentBib}

\bibitem [\protect \citeauthoryear {%
Diniz%
, Siqueira%
\BCBL {}\ \BBA {} Heck%
}{%
Diniz%
\ \protect \BOthers {.}}{%
{\protect \APACyear {2019}}%
}]{%
diniz_taxonomy_2019}
\APACinsertmetastar {%
diniz_taxonomy_2019}%
\begin{APACrefauthors}%
Diniz, E\BPBI H.%
, Siqueira, E\BPBI S.%
\BCBL {}\ \BBA {} Heck, E\BPBI v.%
\end{APACrefauthors}%
\unskip\
\newblock
\APACrefYearMonthDay{2019}{{\APACmonth{01}}}{}.
\newblock
{\BBOQ}\APACrefatitle {Taxonomy of digital community currency platforms}
  {Taxonomy of digital community currency platforms}.{\BBCQ}
\newblock
\APACjournalVolNumPages{Information Technology for Development}{25}{1}{69--91}.
\PrintBackRefs{\CurrentBib}

\bibitem [\protect \citeauthoryear {%
Dye%
, Schatz%
, Rosenberg%
\BCBL {}\ \BBA {} Coleman%
}{%
Dye%
\ \protect \BOthers {.}}{%
{\protect \APACyear {2000}}%
}]{%
dye2000constant}
\APACinsertmetastar {%
dye2000constant}%
\begin{APACrefauthors}%
Dye, J\BPBI F.%
, Schatz, I\BPBI M.%
, Rosenberg, B\BPBI A.%
\BCBL {}\ \BBA {} Coleman, S\BPBI T.%
\end{APACrefauthors}%
\unskip\
\newblock
\APACrefYearMonthDay{2000}{}{}.
\newblock
{\BBOQ}\APACrefatitle {Constant Comparison Method: A Kaleidoscope of Data}
  {Constant comparison method: A kaleidoscope of data}.{\BBCQ}
\newblock
\APACjournalVolNumPages{The Qualitative Report}{4}{1}{1--10}.
\PrintBackRefs{\CurrentBib}

\bibitem [\protect \citeauthoryear {%
{European Commission}%
}{%
{European Commission}%
}{%
{\protect \APACyear {2008}}%
}]{%
european_commission_nace_2008}
\APACinsertmetastar {%
european_commission_nace_2008}%
\begin{APACrefauthors}%
{European Commission}.%
\end{APACrefauthors}%
\unskip\
\newblock
\APACrefYearMonthDay{2008}{}{}.
\newblock
\APACrefbtitle {{NACE} {Rev}. 2 - {Statistical} classification of economic
  activities in the {European} {Community}.} {{NACE} {Rev}. 2 - {Statistical}
  classification of economic activities in the {European} {Community}.}
\newblock
\APACaddressPublisher{}{Eurostat}.
\PrintBackRefs{\CurrentBib}

\bibitem [\protect \citeauthoryear {%
Fellmann%
, Robert%
, Büttner%
, Mucha%
\BCBL {}\ \BBA {} Röcker%
}{%
Fellmann%
\ \protect \BOthers {.}}{%
{\protect \APACyear {2017}}%
}]{%
fellmann_towards_2017}
\APACinsertmetastar {%
fellmann_towards_2017}%
\begin{APACrefauthors}%
Fellmann, M.%
, Robert, S.%
, Büttner, S.%
, Mucha, H.%
\BCBL {}\ \BBA {} Röcker, C.%
\end{APACrefauthors}%
\unskip\
\newblock
\APACrefYearMonthDay{2017}{}{}.
\newblock
{\BBOQ}\APACrefatitle {Towards a {Framework} for {Assistance} {Systems} to
  {Support} {Work} {Processes} in {Smart} {Factories}} {Towards a {Framework}
  for {Assistance} {Systems} to {Support} {Work} {Processes} in {Smart}
  {Factories}}.{\BBCQ}
\newblock
\BIn{} \APACrefbtitle {International {Cross}-{Domain} {Conference} for
  {Machine} {Learning} and {Knowledge} {Extraction}} {International
  {Cross}-{Domain} {Conference} for {Machine} {Learning} and {Knowledge}
  {Extraction}}\ (\BPGS\ 59--68).
\newblock
\APACaddressPublisher{Reggio, Italy}{}.
\PrintBackRefs{\CurrentBib}

\bibitem [\protect \citeauthoryear {%
Fenton%
\ \BBA {} Bieman%
}{%
Fenton%
\ \BBA {} Bieman%
}{%
{\protect \APACyear {2014}}%
}]{%
fenton_software_2014}
\APACinsertmetastar {%
fenton_software_2014}%
\begin{APACrefauthors}%
Fenton, N.%
\BCBT {}\ \BBA {} Bieman, J.%
\end{APACrefauthors}%
\unskip\
\newblock
\APACrefYear{2014}.
\newblock
\APACrefbtitle {Software {Metrics}: {A} {Rigorous} and {Practical} {Approach}}
  {Software {Metrics}: {A} {Rigorous} and {Practical} {Approach}}\
  (\PrintOrdinal{3rd edition}\ \BEd).
\newblock
\APACaddressPublisher{Boca Raton}{CRC Press}.
\PrintBackRefs{\CurrentBib}

\bibitem [\protect \citeauthoryear {%
Ferrucci%
, Levas%
, Bagchi%
, Gondek%
\BCBL {}\ \BBA {} Mueller%
}{%
Ferrucci%
\ \protect \BOthers {.}}{%
{\protect \APACyear {2013}}%
}]{%
ferrucci2013watson}
\APACinsertmetastar {%
ferrucci2013watson}%
\begin{APACrefauthors}%
Ferrucci, D.%
, Levas, A.%
, Bagchi, S.%
, Gondek, D.%
\BCBL {}\ \BBA {} Mueller, E\BPBI T.%
\end{APACrefauthors}%
\unskip\
\newblock
\APACrefYearMonthDay{2013}{}{}.
\newblock
{\BBOQ}\APACrefatitle {Watson: beyond jeopardy!} {Watson: beyond
  jeopardy!}{\BBCQ}
\newblock
\APACjournalVolNumPages{Artificial Intelligence}{199}{}{93--105}.
\PrintBackRefs{\CurrentBib}

\bibitem [\protect \citeauthoryear {%
Fteimi%
\ \BBA {} Lehner%
}{%
Fteimi%
\ \BBA {} Lehner%
}{%
{\protect \APACyear {2018}}%
}]{%
fteimi_analysing_2018}
\APACinsertmetastar {%
fteimi_analysing_2018}%
\begin{APACrefauthors}%
Fteimi, N.%
\BCBT {}\ \BBA {} Lehner, F.%
\end{APACrefauthors}%
\unskip\
\newblock
\APACrefYearMonthDay{2018}{{\APACmonth{01}}}{}.
\newblock
{\BBOQ}\APACrefatitle {Analysing and classifying knowledge management
  publications – a proposed classification scheme} {Analysing and classifying
  knowledge management publications – a proposed classification
  scheme}.{\BBCQ}
\newblock
\APACjournalVolNumPages{Journal of Knowledge Management}{22}{7}{1527--1554}.
\PrintBackRefs{\CurrentBib}

\bibitem [\protect \citeauthoryear {%
Fuggetta%
}{%
Fuggetta%
}{%
{\protect \APACyear {2000}}%
}]{%
fuggetta2000software}
\APACinsertmetastar {%
fuggetta2000software}%
\begin{APACrefauthors}%
Fuggetta, A.%
\end{APACrefauthors}%
\unskip\
\newblock
\APACrefYearMonthDay{2000}{}{}.
\newblock
{\BBOQ}\APACrefatitle {Software process: a roadmap} {Software process: a
  roadmap}.{\BBCQ}
\newblock
\BIn{} \APACrefbtitle {Proceedings of the Conference on the Future of Software
  Engineering} {Proceedings of the conference on the future of software
  engineering}\ (\BPGS\ 25--34).
\PrintBackRefs{\CurrentBib}

\bibitem [\protect \citeauthoryear {%
Gangemi%
, Catenacci%
, Ciaramita%
\BCBL {}\ \BBA {} Lehmann%
}{%
Gangemi%
\ \protect \BOthers {.}}{%
{\protect \APACyear {2006}}%
}]{%
gangemi2006modelling}
\APACinsertmetastar {%
gangemi2006modelling}%
\begin{APACrefauthors}%
Gangemi, A.%
, Catenacci, C.%
, Ciaramita, M.%
\BCBL {}\ \BBA {} Lehmann, J.%
\end{APACrefauthors}%
\unskip\
\newblock
\APACrefYearMonthDay{2006}{}{}.
\newblock
{\BBOQ}\APACrefatitle {Modelling ontology evaluation and validation} {Modelling
  ontology evaluation and validation}.{\BBCQ}
\newblock
\BIn{} \APACrefbtitle {European Semantic Web Conference} {European semantic web
  conference}\ (\BPGS\ 140--154).
\PrintBackRefs{\CurrentBib}

\bibitem [\protect \citeauthoryear {%
Gao%
, Thiebes%
\BCBL {}\ \BBA {} Sunyaev%
}{%
Gao%
\ \protect \BOthers {.}}{%
{\protect \APACyear {2018}}%
}]{%
gao_rethinking_2018}
\APACinsertmetastar {%
gao_rethinking_2018}%
\begin{APACrefauthors}%
Gao, F.%
, Thiebes, S.%
\BCBL {}\ \BBA {} Sunyaev, A.%
\end{APACrefauthors}%
\unskip\
\newblock
\APACrefYearMonthDay{2018}{{\APACmonth{07}}}{}.
\newblock
{\BBOQ}\APACrefatitle {Rethinking the {Meaning} of {Cloud} {Computing} for
  {Health} {Care}: {A} {Taxonomic} {Perspective} and {Future} {Research}
  {Directions}} {Rethinking the {Meaning} of {Cloud} {Computing} for {Health}
  {Care}: {A} {Taxonomic} {Perspective} and {Future} {Research}
  {Directions}}.{\BBCQ}
\newblock
\APACjournalVolNumPages{Journal of Medical Internet Research}{20}{7}{}.
\PrintBackRefs{\CurrentBib}

\bibitem [\protect \citeauthoryear {%
Garshol%
}{%
Garshol%
}{%
{\protect \APACyear {2004}}%
}]{%
garshol_metadata_2004}
\APACinsertmetastar {%
garshol_metadata_2004}%
\begin{APACrefauthors}%
Garshol, L\BPBI M.%
\end{APACrefauthors}%
\unskip\
\newblock
\APACrefYearMonthDay{2004}{{\APACmonth{08}}}{}.
\newblock
{\BBOQ}\APACrefatitle {Metadata? {Thesauri}? {Taxonomies}? {Topic} {Maps}!
  {Making} {Sense} of it all} {Metadata? {Thesauri}? {Taxonomies}? {Topic}
  {Maps}! {Making} {Sense} of it all}.{\BBCQ}
\newblock
\APACjournalVolNumPages{Journal of Information Science}{30}{4}{378--391}.
\newblock
\begin{APACrefDOI} 10.1177/0165551504045856 \end{APACrefDOI}
\PrintBackRefs{\CurrentBib}

\bibitem [\protect \citeauthoryear {%
Ge%
\ \BBA {} Gretzel%
}{%
Ge%
\ \BBA {} Gretzel%
}{%
{\protect \APACyear {2018}}%
}]{%
ge_taxonomy_2018}
\APACinsertmetastar {%
ge_taxonomy_2018}%
\begin{APACrefauthors}%
Ge, J.%
\BCBT {}\ \BBA {} Gretzel, U.%
\end{APACrefauthors}%
\unskip\
\newblock
\APACrefYearMonthDay{2018}{{\APACmonth{01}}}{}.
\newblock
{\BBOQ}\APACrefatitle {A taxonomy of value co-creation on {Weibo} – a
  communication perspective} {A taxonomy of value co-creation on {Weibo} – a
  communication perspective}.{\BBCQ}
\newblock
\APACjournalVolNumPages{International Journal of Contemporary Hospitality
  Management}{30}{4}{2075--2092}.
\PrintBackRefs{\CurrentBib}

\bibitem [\protect \citeauthoryear {%
Geiger%
, Rosemann%
, Fielt%
\BCBL {}\ \BBA {} Schader%
}{%
Geiger%
\ \protect \BOthers {.}}{%
{\protect \APACyear {2012}}%
}]{%
geiger_crowdsourcing_2012}
\APACinsertmetastar {%
geiger_crowdsourcing_2012}%
\begin{APACrefauthors}%
Geiger, D.%
, Rosemann, M.%
, Fielt, E.%
\BCBL {}\ \BBA {} Schader, M.%
\end{APACrefauthors}%
\unskip\
\newblock
\APACrefYearMonthDay{2012}{}{}.
\newblock
{\BBOQ}\APACrefatitle {Crowdsourcing information systems - definition typology,
  and design} {Crowdsourcing information systems - definition typology, and
  design}.{\BBCQ}
\newblock
\BIn{} \APACrefbtitle {Proceedings of the 33rd {International} {Conference} on
  {Information} {Systems}.} {Proceedings of the 33rd {International}
  {Conference} on {Information} {Systems}.}
\newblock
\APACaddressPublisher{Atlanta, Georgia, USA}{AIS Electronic Library}.
\PrintBackRefs{\CurrentBib}

\bibitem [\protect \citeauthoryear {%
Gelder%
}{%
Gelder%
}{%
{\protect \APACyear {2015}}%
}]{%
gelder_design_2015}
\APACinsertmetastar {%
gelder_design_2015}%
\begin{APACrefauthors}%
Gelder, J.%
\end{APACrefauthors}%
\unskip\
\newblock
\APACrefYearMonthDay{2015}{{\APACmonth{09}}}{}.
\newblock
{\BBOQ}\APACrefatitle {The design and development of a classification system
  for {BIM}} {The design and development of a classification system for
  {BIM}}.{\BBCQ}
\newblock
\BIn{} (\BVOL~149, \BPGS\ 477--491).
\newblock
\APACaddressPublisher{}{WIT Press}.
\PrintBackRefs{\CurrentBib}

\bibitem [\protect \citeauthoryear {%
Gibbs%
, Gretzel%
\BCBL {}\ \BBA {} Saltzman%
}{%
Gibbs%
\ \protect \BOthers {.}}{%
{\protect \APACyear {2016}}%
}]{%
gibbs_experience-based_2016}
\APACinsertmetastar {%
gibbs_experience-based_2016}%
\begin{APACrefauthors}%
Gibbs, C.%
, Gretzel, U.%
\BCBL {}\ \BBA {} Saltzman, J.%
\end{APACrefauthors}%
\unskip\
\newblock
\APACrefYearMonthDay{2016}{{\APACmonth{06}}}{}.
\newblock
{\BBOQ}\APACrefatitle {An experience-based taxonomy of branded hotel mobile
  application features} {An experience-based taxonomy of branded hotel mobile
  application features}.{\BBCQ}
\newblock
\APACjournalVolNumPages{Information Technology \& Tourism}{16}{2}{175--199}.
\PrintBackRefs{\CurrentBib}

\bibitem [\protect \citeauthoryear {%
Gimpel%
, Rau%
\BCBL {}\ \BBA {} Röglinger%
}{%
Gimpel%
\ \protect \BOthers {.}}{%
{\protect \APACyear {2018}}%
}]{%
gimpel_understanding_2018}
\APACinsertmetastar {%
gimpel_understanding_2018}%
\begin{APACrefauthors}%
Gimpel, H.%
, Rau, D.%
\BCBL {}\ \BBA {} Röglinger, M.%
\end{APACrefauthors}%
\unskip\
\newblock
\APACrefYearMonthDay{2018}{{\APACmonth{08}}}{}.
\newblock
{\BBOQ}\APACrefatitle {Understanding {FinTech} start-ups – a taxonomy of
  consumer-oriented service offerings} {Understanding {FinTech} start-ups – a
  taxonomy of consumer-oriented service offerings}.{\BBCQ}
\newblock
\APACjournalVolNumPages{Electronic Markets}{28}{3}{245--264}.
\PrintBackRefs{\CurrentBib}

\bibitem [\protect \citeauthoryear {%
Gregor%
}{%
Gregor%
}{%
{\protect \APACyear {2006}}%
}]{%
gregor_nature_2006}
\APACinsertmetastar {%
gregor_nature_2006}%
\begin{APACrefauthors}%
Gregor, S.%
\end{APACrefauthors}%
\unskip\
\newblock
\APACrefYearMonthDay{2006}{}{}.
\newblock
{\BBOQ}\APACrefatitle {The {Nature} of {Theory} in {Information} {Systems}}
  {The {Nature} of {Theory} in {Information} {Systems}}.{\BBCQ}
\newblock
\APACjournalVolNumPages{MIS Quarterly}{30}{3}{611--642}.
\PrintBackRefs{\CurrentBib}

\bibitem [\protect \citeauthoryear {%
Gruninger%
, Bodenreider%
, Olken%
, Obrst%
\BCBL {}\ \BBA {} Yim%
}{%
Gruninger%
\ \protect \BOthers {.}}{%
{\protect \APACyear {2008}}%
}]{%
gruninger_ontology_2008}
\APACinsertmetastar {%
gruninger_ontology_2008}%
\begin{APACrefauthors}%
Gruninger, M.%
, Bodenreider, O.%
, Olken, F.%
, Obrst, L.%
\BCBL {}\ \BBA {} Yim, P.%
\end{APACrefauthors}%
\unskip\
\newblock
\APACrefYearMonthDay{2008}{}{}.
\newblock
{\BBOQ}\APACrefatitle {Ontology {Summit} 2007–{Ontology}, taxonomy,
  folksonomy: {Understanding} the distinctions} {Ontology {Summit}
  2007–{Ontology}, taxonomy, folksonomy: {Understanding} the
  distinctions}.{\BBCQ}
\newblock
\APACjournalVolNumPages{Applied Ontology}{3}{3}{191--200}.
\PrintBackRefs{\CurrentBib}

\bibitem [\protect \citeauthoryear {%
Gwet%
}{%
Gwet%
}{%
{\protect \APACyear {2014}}%
}]{%
gwet_handbook_2014}
\APACinsertmetastar {%
gwet_handbook_2014}%
\begin{APACrefauthors}%
Gwet, K\BPBI L.%
\end{APACrefauthors}%
\unskip\
\newblock
\APACrefYear{2014}.
\newblock
\APACrefbtitle {Handbook of inter-rater reliability: the definitive guide to
  measuring the extent of agreement among raters} {Handbook of inter-rater
  reliability: the definitive guide to measuring the extent of agreement among
  raters}\ (\PrintOrdinal{4th}\ \BEd).
\newblock
\APACaddressPublisher{}{Advanced Analytics, LLC}.
\PrintBackRefs{\CurrentBib}

\bibitem [\protect \citeauthoryear {%
Herterich%
, Holler%
, Uebernickel%
\BCBL {}\ \BBA {} Brenner%
}{%
Herterich%
\ \protect \BOthers {.}}{%
{\protect \APACyear {2015}}%
}]{%
herterich_understanding_2015}
\APACinsertmetastar {%
herterich_understanding_2015}%
\begin{APACrefauthors}%
Herterich, M.%
, Holler, M.%
, Uebernickel, F.%
\BCBL {}\ \BBA {} Brenner, W.%
\end{APACrefauthors}%
\unskip\
\newblock
\APACrefYearMonthDay{2015}{{\APACmonth{05}}}{}.
\newblock
{\BBOQ}\APACrefatitle {Understanding the {Business} {Value}: {Towards} a
  {Taxonomy} of {Industrial} {Use} {Scenarios} enabled by {Cyber}-{Physical}
  {Systems} in the {Equipment} {Manufacturing} {Industry}} {Understanding the
  {Business} {Value}: {Towards} a {Taxonomy} of {Industrial} {Use} {Scenarios}
  enabled by {Cyber}-{Physical} {Systems} in the {Equipment} {Manufacturing}
  {Industry}}.{\BBCQ}
\newblock
\BIn{} \APACrefbtitle {Proceedings {Internation} {Conference} on {Information}
  {Resources} {Management}.} {Proceedings {Internation} {Conference} on
  {Information} {Resources} {Management}.}
\newblock
\APACaddressPublisher{Ottawa, Canada}{AIS Electronic Library}.
\PrintBackRefs{\CurrentBib}

\bibitem [\protect \citeauthoryear {%
Herzfeldt%
, Hausen%
, Briggs%
\BCBL {}\ \BBA {} Krcmar%
}{%
Herzfeldt%
\ \protect \BOthers {.}}{%
{\protect \APACyear {2012}}%
}]{%
herzfeldt_developin_2012}
\APACinsertmetastar {%
herzfeldt_developin_2012}%
\begin{APACrefauthors}%
Herzfeldt, A.%
, Hausen, M.%
, Briggs, R\BPBI O.%
\BCBL {}\ \BBA {} Krcmar, H.%
\end{APACrefauthors}%
\unskip\
\newblock
\APACrefYearMonthDay{2012}{{\APACmonth{06}}}{}.
\newblock
{\BBOQ}\APACrefatitle {Developin a {Risk} {Management} {Process} and {Risk}
  {Taxonomy} for medium-sized {IT} {Solution} {Providers}} {Developin a {Risk}
  {Management} {Process} and {Risk} {Taxonomy} for medium-sized {IT} {Solution}
  {Providers}}.{\BBCQ}
\newblock
\BIn{} \APACrefbtitle {Proceedings {European} {Conference} on {Information}
  {Systems}.} {Proceedings {European} {Conference} on {Information} {Systems}.}
\newblock
\APACaddressPublisher{Barcelona, Spain}{AIS Electronic Library}.
\PrintBackRefs{\CurrentBib}

\bibitem [\protect \citeauthoryear {%
Hlomani%
\ \BBA {} Stacey%
}{%
Hlomani%
\ \BBA {} Stacey%
}{%
{\protect \APACyear {2014}}%
}]{%
hlomani2014approaches}
\APACinsertmetastar {%
hlomani2014approaches}%
\begin{APACrefauthors}%
Hlomani, H.%
\BCBT {}\ \BBA {} Stacey, D.%
\end{APACrefauthors}%
\unskip\
\newblock
\APACrefYearMonthDay{2014}{}{}.
\newblock
{\BBOQ}\APACrefatitle {Approaches, methods, metrics, measures, and subjectivity
  in ontology evaluation: A survey} {Approaches, methods, metrics, measures,
  and subjectivity in ontology evaluation: A survey}.{\BBCQ}
\newblock
\APACjournalVolNumPages{Semantic Web Journal}{1}{5}{1--11}.
\PrintBackRefs{\CurrentBib}

\bibitem [\protect \citeauthoryear {%
Holler%
, Uebernickel%
\BCBL {}\ \BBA {} Brenner%
}{%
Holler%
\ \protect \BOthers {.}}{%
{\protect \APACyear {2017}}%
}]{%
holler_defining_2017}
\APACinsertmetastar {%
holler_defining_2017}%
\begin{APACrefauthors}%
Holler, M.%
, Uebernickel, F.%
\BCBL {}\ \BBA {} Brenner, W.%
\end{APACrefauthors}%
\unskip\
\newblock
\APACrefYearMonthDay{2017}{{\APACmonth{06}}}{}.
\newblock
{\BBOQ}\APACrefatitle {Defining {Archetypes} of {E}-{Collaboration} for
  {Product} {Development} in {The} {Automotive} {Industry}} {Defining
  {Archetypes} of {E}-{Collaboration} for {Product} {Development} in {The}
  {Automotive} {Industry}}.{\BBCQ}
\newblock
\BIn{} \APACrefbtitle {Proceedings {European} {Conference} on {Information}
  {Systems}.} {Proceedings {European} {Conference} on {Information} {Systems}.}
\newblock
\APACaddressPublisher{Guimarães, Portugal}{AIS Electronic Library}.
\PrintBackRefs{\CurrentBib}

\bibitem [\protect \citeauthoryear {%
Institute%
}{%
Institute%
}{%
{\protect \APACyear {2019}}%
}]{%
the_construction_specifications_institute_omniclass_2019}
\APACinsertmetastar {%
the_construction_specifications_institute_omniclass_2019}%
\begin{APACrefauthors}%
Institute, T\BPBI C\BPBI S.%
\end{APACrefauthors}%
\unskip\
\newblock
\APACrefYearMonthDay{2019}{}{}.
\newblock
\APACrefbtitle {{OmniClass} - {A} {Strategy} for {Classifying} the {Built}
  {Environment}.} {{OmniClass} - {A} {Strategy} for {Classifying} the {Built}
  {Environment}.}
\PrintBackRefs{\CurrentBib}

\bibitem [\protect \citeauthoryear {%
Jarvinen%
}{%
Jarvinen%
}{%
{\protect \APACyear {2000}}%
}]{%
jarvinen_research_2000}
\APACinsertmetastar {%
jarvinen_research_2000}%
\begin{APACrefauthors}%
Jarvinen, P.%
\end{APACrefauthors}%
\unskip\
\newblock
\APACrefYearMonthDay{2000}{{\APACmonth{07}}}{}.
\newblock
{\BBOQ}\APACrefatitle {Research {Questions} {Guiding} {Selection} of an
  {Appropriate} {Research} {Method}} {Research {Questions} {Guiding}
  {Selection} of an {Appropriate} {Research} {Method}}.{\BBCQ}
\newblock
\BIn{} \APACrefbtitle {Proceedings {European} {Conference} on {Information}
  {Systems}.} {Proceedings {European} {Conference} on {Information} {Systems}.}
\newblock
\APACaddressPublisher{Vienna, Austria}{AIS Electronic Library}.
\PrintBackRefs{\CurrentBib}

\bibitem [\protect \citeauthoryear {%
Jöhnk%
, Röglinger%
, Thimmel%
\BCBL {}\ \BBA {} Urbach%
}{%
Jöhnk%
\ \protect \BOthers {.}}{%
{\protect \APACyear {0217}}%
}]{%
johnk_how_0217}
\APACinsertmetastar {%
johnk_how_0217}%
\begin{APACrefauthors}%
Jöhnk, J.%
, Röglinger, M.%
, Thimmel, M.%
\BCBL {}\ \BBA {} Urbach, N.%
\end{APACrefauthors}%
\unskip\
\newblock
\APACrefYearMonthDay{0217}{{\APACmonth{06}}}{}.
\newblock
{\BBOQ}\APACrefatitle {How to implement {Agile} {IT} setups: {A} {Taxonomy} of
  {Design} {Options}} {How to implement {Agile} {IT} setups: {A} {Taxonomy} of
  {Design} {Options}}.{\BBCQ}
\newblock
\BIn{} \APACrefbtitle {Proceedings {European} {Conference} on {Information}
  {Systems}} {Proceedings {European} {Conference} on {Information} {Systems}}\
  (\BPGS\ 1521--1535).
\newblock
\APACaddressPublisher{Guimarães, Portugal}{AIS Electronic Library}.
\PrintBackRefs{\CurrentBib}

\bibitem [\protect \citeauthoryear {%
Keller%
\ \BBA {} König%
}{%
Keller%
\ \BBA {} König%
}{%
{\protect \APACyear {2014}}%
}]{%
keller_reference_2014}
\APACinsertmetastar {%
keller_reference_2014}%
\begin{APACrefauthors}%
Keller, R.%
\BCBT {}\ \BBA {} König, C.%
\end{APACrefauthors}%
\unskip\
\newblock
\APACrefYearMonthDay{2014}{{\APACmonth{12}}}{}.
\newblock
{\BBOQ}\APACrefatitle {A {Reference} {Model} to {Support} {Risk}
  {Identification} in {Cloud} {Networks}} {A {Reference} {Model} to {Support}
  {Risk} {Identification} in {Cloud} {Networks}}.{\BBCQ}.
\PrintBackRefs{\CurrentBib}

\bibitem [\protect \citeauthoryear {%
Khosravi%
\ \BBA {} Gueheneuc%
}{%
Khosravi%
\ \BBA {} Gueheneuc%
}{%
{\protect \APACyear {2004}}%
}]{%
khosravi_quality_2004}
\APACinsertmetastar {%
khosravi_quality_2004}%
\begin{APACrefauthors}%
Khosravi, K.%
\BCBT {}\ \BBA {} Gueheneuc, Y\BHBI G.%
\end{APACrefauthors}%
\unskip\
\newblock
\APACrefYearMonthDay{2004}{}{}.
\newblock
{\BBOQ}\APACrefatitle {A Quality Model for Design Patterns} {A quality model
  for design patterns}.{\BBCQ}
\newblock
\APACjournalVolNumPages{German Industry Standard}{}{}{108}.
\PrintBackRefs{\CurrentBib}

\bibitem [\protect \citeauthoryear {%
King%
\ \BBA {} Jr%
}{%
King%
\ \BBA {} Jr%
}{%
{\protect \APACyear {1999}}%
}]{%
king_empirical_1999}
\APACinsertmetastar {%
king_empirical_1999}%
\begin{APACrefauthors}%
King, W\BPBI R.%
\BCBT {}\ \BBA {} Jr, V\BPBI S.%
\end{APACrefauthors}%
\unskip\
\newblock
\APACrefYearMonthDay{1999}{{\APACmonth{03}}}{}.
\newblock
{\BBOQ}\APACrefatitle {An {Empirical} {Assessment} of the {Organization} of
  {Transnational} {Information} {Systems}} {An {Empirical} {Assessment} of the
  {Organization} of {Transnational} {Information} {Systems}}.{\BBCQ}
\newblock
\APACjournalVolNumPages{Journal of Management Information
  Systems}{15}{4}{7--28}.
\PrintBackRefs{\CurrentBib}

\bibitem [\protect \citeauthoryear {%
Küpper%
, Jung%
, Lehmkuhl%
\BCBL {}\ \BBA {} Wieneke%
}{%
Küpper%
\ \protect \BOthers {.}}{%
{\protect \APACyear {2014}}%
}]{%
kupper_features_2014}
\APACinsertmetastar {%
kupper_features_2014}%
\begin{APACrefauthors}%
Küpper, T.%
, Jung, R.%
, Lehmkuhl, T.%
\BCBL {}\ \BBA {} Wieneke, A.%
\end{APACrefauthors}%
\unskip\
\newblock
\APACrefYearMonthDay{2014}{{\APACmonth{08}}}{}.
\newblock
{\BBOQ}\APACrefatitle {Features for {Social} {CRM} {Technology} - {An}
  {Organizational} {Perspective}} {Features for {Social} {CRM} {Technology} -
  {An} {Organizational} {Perspective}}.{\BBCQ}
\newblock
\BIn{} \APACrefbtitle {Proceedings of the 20st {Americas} {Conference} on
  {Information} {Systems} ({AMCIS} 2014)} {Proceedings of the 20st {Americas}
  {Conference} on {Information} {Systems} ({AMCIS} 2014)}\ (\BPGS\ 1--10).
\newblock
\APACaddressPublisher{Savannah, Georgia, USA}{AIS Electronic Library}.
\newblock
\APACrefnote{Num Pages: 10}
\PrintBackRefs{\CurrentBib}

\bibitem [\protect \citeauthoryear {%
Labazova%
, Dehling%
\BCBL {}\ \BBA {} Sunyaev%
}{%
Labazova%
\ \protect \BOthers {.}}{%
{\protect \APACyear {2018}}%
}]{%
labazova_hype_2018}
\APACinsertmetastar {%
labazova_hype_2018}%
\begin{APACrefauthors}%
Labazova, O.%
, Dehling, T.%
\BCBL {}\ \BBA {} Sunyaev, A.%
\end{APACrefauthors}%
\unskip\
\newblock
\APACrefYearMonthDay{2018}{}{}.
\newblock
{\BBOQ}\APACrefatitle {From {Hype} to {Reality}: {A} {Taxonomy} of {Blockchain}
  {Applications}} {From {Hype} to {Reality}: {A} {Taxonomy} of {Blockchain}
  {Applications}}.{\BBCQ}
\newblock
\BIn{} \APACrefbtitle {Proceedings of the 52nd {Hawaii} {International}
  {Conference} on {System} {Sciences} ({HICSS} 2019).} {Proceedings of the 52nd
  {Hawaii} {International} {Conference} on {System} {Sciences} ({HICSS} 2019).}
\newblock
\APACaddressPublisher{Wailea, Maui, HI, USA}{}.
\PrintBackRefs{\CurrentBib}

\bibitem [\protect \citeauthoryear {%
Larsen%
}{%
Larsen%
}{%
{\protect \APACyear {2003}}%
}]{%
larsen_taxonomy_2003}
\APACinsertmetastar {%
larsen_taxonomy_2003}%
\begin{APACrefauthors}%
Larsen, K\BPBI R\BPBI T.%
\end{APACrefauthors}%
\unskip\
\newblock
\APACrefYearMonthDay{2003}{{\APACmonth{10}}}{}.
\newblock
{\BBOQ}\APACrefatitle {A {Taxonomy} of {Antecedents} of {Information} {Systems}
  {Success}: {Variable} {Analysis} {Studies}} {A {Taxonomy} of {Antecedents} of
  {Information} {Systems} {Success}: {Variable} {Analysis} {Studies}}.{\BBCQ}
\newblock
\APACjournalVolNumPages{Journal of Management Information
  Systems}{20}{2}{169--246}.
\PrintBackRefs{\CurrentBib}

\bibitem [\protect \citeauthoryear {%
Li%
\ \BBA {} Cleland-Huang%
}{%
Li%
\ \BBA {} Cleland-Huang%
}{%
{\protect \APACyear {2013}}%
}]{%
li_ontology-based_2013}
\APACinsertmetastar {%
li_ontology-based_2013}%
\begin{APACrefauthors}%
Li, Y.%
\BCBT {}\ \BBA {} Cleland-Huang, J.%
\end{APACrefauthors}%
\unskip\
\newblock
\APACrefYearMonthDay{2013}{{\APACmonth{05}}}{}.
\newblock
{\BBOQ}\APACrefatitle {Ontology-based trace retrieval} {Ontology-based trace
  retrieval}.{\BBCQ}
\newblock
\BIn{} \APACrefbtitle {2013 7th {International} {Workshop} on {Traceability} in
  {Emerging} {Forms} of {Software} {Engineering} ({TEFSE})} {2013 7th
  {International} {Workshop} on {Traceability} in {Emerging} {Forms} of
  {Software} {Engineering} ({TEFSE})}\ (\BPGS\ 30--36).
\newblock
\APACaddressPublisher{San Francisco, USA}{IEEE}.
\PrintBackRefs{\CurrentBib}

\bibitem [\protect \citeauthoryear {%
Mahaini%
, Li%
\BCBL {}\ \BBA {} Sağlam%
}{%
Mahaini%
\ \protect \BOthers {.}}{%
{\protect \APACyear {2019}}%
}]{%
mahaini_building_2019}
\APACinsertmetastar {%
mahaini_building_2019}%
\begin{APACrefauthors}%
Mahaini, M\BPBI I.%
, Li, S.%
\BCBL {}\ \BBA {} Sağlam, R\BPBI B.%
\end{APACrefauthors}%
\unskip\
\newblock
\APACrefYearMonthDay{2019}{{\APACmonth{08}}}{}.
\newblock
{\BBOQ}\APACrefatitle {Building {Taxonomies} based on {Human}-{Machine}
  {Teaming}: {Cyber} {Security} as an {Example}} {Building {Taxonomies} based
  on {Human}-{Machine} {Teaming}: {Cyber} {Security} as an {Example}}.{\BBCQ}
\newblock
\BIn{} \APACrefbtitle {Proceedings of the 14th {International} {Conference} on
  {Availability}, {Reliability} and {Security}} {Proceedings of the 14th
  {International} {Conference} on {Availability}, {Reliability} and
  {Security}}\ (\BPGS\ 1--9).
\newblock
\APACaddressPublisher{Canterbury, UK}{ACM}.
\PrintBackRefs{\CurrentBib}

\bibitem [\protect \citeauthoryear {%
McDaniel%
\ \BBA {} Storey%
}{%
McDaniel%
\ \BBA {} Storey%
}{%
{\protect \APACyear {2019}}%
}]{%
mcdaniel2019evaluating}
\APACinsertmetastar {%
mcdaniel2019evaluating}%
\begin{APACrefauthors}%
McDaniel, M.%
\BCBT {}\ \BBA {} Storey, V\BPBI C.%
\end{APACrefauthors}%
\unskip\
\newblock
\APACrefYearMonthDay{2019}{}{}.
\newblock
{\BBOQ}\APACrefatitle {Evaluating domain ontologies: clarification,
  classification, and challenges} {Evaluating domain ontologies: clarification,
  classification, and challenges}.{\BBCQ}
\newblock
\APACjournalVolNumPages{ACM Computing Surveys (CSUR)}{52}{4}{1--44}.
\PrintBackRefs{\CurrentBib}

\bibitem [\protect \citeauthoryear {%
Mikolov%
, Sutskever%
, Chen%
, Corrado%
\BCBL {}\ \BBA {} Dean%
}{%
Mikolov%
\ \protect \BOthers {.}}{%
{\protect \APACyear {2013}}%
}]{%
mikolov_distributed_2013}
\APACinsertmetastar {%
mikolov_distributed_2013}%
\begin{APACrefauthors}%
Mikolov, T.%
, Sutskever, I.%
, Chen, K.%
, Corrado, G.%
\BCBL {}\ \BBA {} Dean, J.%
\end{APACrefauthors}%
\unskip\
\newblock
\APACrefYearMonthDay{2013}{{\APACmonth{10}}}{}.
\newblock
{\BBOQ}\APACrefatitle {Distributed {Representations} of {Words} and {Phrases}
  and their {Compositionality}} {Distributed {Representations} of {Words} and
  {Phrases} and their {Compositionality}}.{\BBCQ}
\newblock
\APACjournalVolNumPages{arXiv:1310.4546 [cs, stat]}{}{}{}.
\newblock
\begin{APACrefURL} [{2021-05-03}]\url{http://arxiv.org/abs/1310.4546}
  \end{APACrefURL}
\newblock
\APACrefnote{arXiv: 1310.4546}
\PrintBackRefs{\CurrentBib}

\bibitem [\protect \citeauthoryear {%
Miller%
}{%
Miller%
}{%
{\protect \APACyear {1956}}%
}]{%
miller_magical_1956}
\APACinsertmetastar {%
miller_magical_1956}%
\begin{APACrefauthors}%
Miller, G\BPBI A.%
\end{APACrefauthors}%
\unskip\
\newblock
\APACrefYearMonthDay{1956}{}{}.
\newblock
{\BBOQ}\APACrefatitle {The {Magical} {Number} {Seven}, {Plus} or {Minus} {Two}:
  {Some} {Limits} on {Our} {Capacity} for {Processing} {Information}} {The
  {Magical} {Number} {Seven}, {Plus} or {Minus} {Two}: {Some} {Limits} on {Our}
  {Capacity} for {Processing} {Information}}.{\BBCQ}
\newblock
\APACjournalVolNumPages{Psychological Review}{63}{2}{81--97}.
\PrintBackRefs{\CurrentBib}

\bibitem [\protect \citeauthoryear {%
Nai~Fovino%
\ \protect \BOthers {.}}{%
Nai~Fovino%
\ \protect \BOthers {.}}{%
{\protect \APACyear {2019}}%
}]{%
nai_fovino_proposal_2019}
\APACinsertmetastar {%
nai_fovino_proposal_2019}%
\begin{APACrefauthors}%
Nai~Fovino, I.%
, Neisse, R.%
, Hernandez~Ramos, J.%
, Polemi, N.%
, Ruzzante, G.%
, Figwer, M.%
\BCBL {}\ \BBA {} Lazari, A.%
\end{APACrefauthors}%
\unskip\
\newblock
\APACrefYearMonthDay{2019}{}{}.
\newblock
\APACrefbtitle {A proposal for a {European} cybersecurity taxonomy.} {A
  proposal for a {European} cybersecurity taxonomy.}\ \APACbVolEdTR{}{\BTR{}\
  \BNUM\ EUR 29868 EN}.
\newblock
\APACaddressInstitution{Luxembourg}{Publications Office of the European Union}.
\newblock
\begin{APACrefURL}
  [{2021-05-13}]\url{https://data.europa.eu/doi/10.2760/106002}
  \end{APACrefURL}
\PrintBackRefs{\CurrentBib}

\bibitem [\protect \citeauthoryear {%
{NBS}%
}{%
{NBS}%
}{%
{\protect \APACyear {2021}}%
{\protect \APACexlab {{\protect \BCnt {1}}}}}]{%
nbs_uniclass_2021}
\APACinsertmetastar {%
nbs_uniclass_2021}%
\begin{APACrefauthors}%
{NBS}.%
\end{APACrefauthors}%
\unskip\
\newblock
\APACrefYearMonthDay{2021{\protect \BCnt {1}}}{}{}.
\newblock
\APACrefbtitle {Uniclass 2015 - {January} 2021 update.} {Uniclass 2015 -
  {January} 2021 update.}
\newblock
\begin{APACrefURL}
  [{2021-05-12}]\url{https://www.thenbs.com/knowledge/uniclass-2015-january-2021-update}
  \end{APACrefURL}
\PrintBackRefs{\CurrentBib}

\bibitem [\protect \citeauthoryear {%
{NBS}%
}{%
{NBS}%
}{%
{\protect \APACyear {2021}}%
{\protect \APACexlab {{\protect \BCnt {2}}}}}]{%
nbs_what_2021}
\APACinsertmetastar {%
nbs_what_2021}%
\begin{APACrefauthors}%
{NBS}.%
\end{APACrefauthors}%
\unskip\
\newblock
\APACrefYearMonthDay{2021{\protect \BCnt {2}}}{}{}.
\newblock
\APACrefbtitle {What is {Uniclass} 2015?} {What is {Uniclass} 2015?}
\newblock
\begin{APACrefURL}
  [{2021-05-13}]\url{https://www.thenbs.com/knowledge/what-is-uniclass-2015}
  \end{APACrefURL}
\PrintBackRefs{\CurrentBib}

\bibitem [\protect \citeauthoryear {%
Nickerson%
, Varshney%
\BCBL {}\ \BBA {} Muntermann%
}{%
Nickerson%
\ \protect \BOthers {.}}{%
{\protect \APACyear {2013}}%
}]{%
nickerson_method_2013}
\APACinsertmetastar {%
nickerson_method_2013}%
\begin{APACrefauthors}%
Nickerson, R\BPBI C.%
, Varshney, U.%
\BCBL {}\ \BBA {} Muntermann, J.%
\end{APACrefauthors}%
\unskip\
\newblock
\APACrefYearMonthDay{2013}{{\APACmonth{05}}}{}.
\newblock
{\BBOQ}\APACrefatitle {A method for taxonomy development and its application in
  information systems} {A method for taxonomy development and its application
  in information systems}.{\BBCQ}
\newblock
\APACjournalVolNumPages{European Journal of Information
  Systems}{22}{3}{336--359}.
\PrintBackRefs{\CurrentBib}

\bibitem [\protect \citeauthoryear {%
Nooralahzadeh%
, Øvrelid%
\BCBL {}\ \BBA {} Lønning%
}{%
Nooralahzadeh%
\ \protect \BOthers {.}}{%
{\protect \APACyear {2018}}%
}]{%
nooralahzadeh_evaluation_2018}
\APACinsertmetastar {%
nooralahzadeh_evaluation_2018}%
\begin{APACrefauthors}%
Nooralahzadeh, F.%
, Øvrelid, L.%
\BCBL {}\ \BBA {} Lønning, J\BPBI T.%
\end{APACrefauthors}%
\unskip\
\newblock
\APACrefYearMonthDay{2018}{{\APACmonth{05}}}{}.
\newblock
{\BBOQ}\APACrefatitle {Evaluation of {Domain}-specific {Word} {Embeddings}
  using {Knowledge} {Resources}} {Evaluation of {Domain}-specific {Word}
  {Embeddings} using {Knowledge} {Resources}}.{\BBCQ}
\newblock
\BIn{} \APACrefbtitle {Proceedings of the {Eleventh} {International}
  {Conference} on {Language} {Resources} and {Evaluation} ({LREC} 2018).}
  {Proceedings of the {Eleventh} {International} {Conference} on {Language}
  {Resources} and {Evaluation} ({LREC} 2018).}
\newblock
\APACaddressPublisher{Miyazaki, Japan}{European Language Resources Association
  (ELRA)}.
\PrintBackRefs{\CurrentBib}

\bibitem [\protect \citeauthoryear {%
Oberländer%
, Röglinger%
, Rosemann%
\BCBL {}\ \BBA {} Kees%
}{%
Oberländer%
\ \protect \BOthers {.}}{%
{\protect \APACyear {2018}}%
}]{%
oberlander_conceptualizing_2018}
\APACinsertmetastar {%
oberlander_conceptualizing_2018}%
\begin{APACrefauthors}%
Oberländer, A\BPBI M.%
, Röglinger, M.%
, Rosemann, M.%
\BCBL {}\ \BBA {} Kees, A.%
\end{APACrefauthors}%
\unskip\
\newblock
\APACrefYearMonthDay{2018}{{\APACmonth{07}}}{}.
\newblock
{\BBOQ}\APACrefatitle {Conceptualizing business-to-thing interactions – {A}
  sociomaterial perspective on the {Internet} of {Things}} {Conceptualizing
  business-to-thing interactions – {A} sociomaterial perspective on the
  {Internet} of {Things}}.{\BBCQ}
\newblock
\APACjournalVolNumPages{European Journal of Information
  Systems}{27}{4}{486--502}.
\PrintBackRefs{\CurrentBib}

\bibitem [\protect \citeauthoryear {%
Obrst%
, Ceusters%
, Mani%
, Ray%
\BCBL {}\ \BBA {} Smith%
}{%
Obrst%
\ \protect \BOthers {.}}{%
{\protect \APACyear {2007}}%
}]{%
obrst2007evaluation}
\APACinsertmetastar {%
obrst2007evaluation}%
\begin{APACrefauthors}%
Obrst, L.%
, Ceusters, W.%
, Mani, I.%
, Ray, S.%
\BCBL {}\ \BBA {} Smith, B.%
\end{APACrefauthors}%
\unskip\
\newblock
\APACrefYearMonthDay{2007}{}{}.
\newblock
{\BBOQ}\APACrefatitle {The evaluation of ontologies} {The evaluation of
  ontologies}.{\BBCQ}
\newblock
\BIn{} \APACrefbtitle {Semantic web} {Semantic web}\ (\BPGS\ 139--158).
\newblock
\APACaddressPublisher{}{Springer}.
\PrintBackRefs{\CurrentBib}

\bibitem [\protect \citeauthoryear {%
of~the President Office~of Management {and}~Budget%
}{%
of~the President Office~of Management {and}~Budget%
}{%
{\protect \APACyear {2017}}%
}]{%
executive_office_of_the_president_office_of_management_and_budget_north_2017}
\APACinsertmetastar {%
executive_office_of_the_president_office_of_management_and_budget_north_2017}%
\begin{APACrefauthors}%
of~the President Office~of Management {and}~Budget, E\BPBI O.%
\end{APACrefauthors}%
\unskip\
\newblock
\APACrefYearMonthDay{2017}{}{}.
\newblock
\APACrefbtitle {North {American} {Industry} {Classification} {System}.} {North
  {American} {Industry} {Classification} {System}.}
\newblock
\APACaddressPublisher{}{United States Government}.
\PrintBackRefs{\CurrentBib}

\bibitem [\protect \citeauthoryear {%
Panichella%
\ \protect \BOthers {.}}{%
Panichella%
\ \protect \BOthers {.}}{%
{\protect \APACyear {2015}}%
}]{%
panichella_how_2015}
\APACinsertmetastar {%
panichella_how_2015}%
\begin{APACrefauthors}%
Panichella, S.%
, Di~Sorbo, A.%
, Guzman, E.%
, Visaggio, C\BPBI A.%
, Canfora, G.%
\BCBL {}\ \BBA {} Gall, H\BPBI C.%
\end{APACrefauthors}%
\unskip\
\newblock
\APACrefYearMonthDay{2015}{{\APACmonth{09}}}{}.
\newblock
{\BBOQ}\APACrefatitle {How can i improve my app? {Classifying} user reviews for
  software maintenance and evolution} {How can i improve my app? {Classifying}
  user reviews for software maintenance and evolution}.{\BBCQ}
\newblock
\BIn{} \APACrefbtitle {2015 {IEEE} {International} {Conference} on {Software}
  {Maintenance} and {Evolution} ({ICSME})} {2015 {IEEE} {International}
  {Conference} on {Software} {Maintenance} and {Evolution} ({ICSME})}\ (\BPGS\
  281--290).
\newblock
\APACaddressPublisher{Bremen, Germany}{}.
\PrintBackRefs{\CurrentBib}

\bibitem [\protect \citeauthoryear {%
Prat%
, Comyn-Wattiau%
\BCBL {}\ \BBA {} Akoka%
}{%
Prat%
\ \protect \BOthers {.}}{%
{\protect \APACyear {2015}}%
}]{%
prat_taxonomy_2015}
\APACinsertmetastar {%
prat_taxonomy_2015}%
\begin{APACrefauthors}%
Prat, N.%
, Comyn-Wattiau, I.%
\BCBL {}\ \BBA {} Akoka, J.%
\end{APACrefauthors}%
\unskip\
\newblock
\APACrefYearMonthDay{2015}{{\APACmonth{07}}}{}.
\newblock
{\BBOQ}\APACrefatitle {A {Taxonomy} of {Evaluation} {Methods} for {Information}
  {Systems} {Artifacts}} {A {Taxonomy} of {Evaluation} {Methods} for
  {Information} {Systems} {Artifacts}}.{\BBCQ}
\newblock
\APACjournalVolNumPages{Journal of Management Information
  Systems}{32}{3}{229--267}.
\PrintBackRefs{\CurrentBib}

\bibitem [\protect \citeauthoryear {%
Püschel%
, Schlott%
\BCBL {}\ \BBA {} Röglinger%
}{%
Püschel%
\ \protect \BOthers {.}}{%
{\protect \APACyear {2016}}%
}]{%
puschel_whats_2016}
\APACinsertmetastar {%
puschel_whats_2016}%
\begin{APACrefauthors}%
Püschel, L.%
, Schlott, H.%
\BCBL {}\ \BBA {} Röglinger, M.%
\end{APACrefauthors}%
\unskip\
\newblock
\APACrefYearMonthDay{2016}{{\APACmonth{12}}}{}.
\newblock
{\BBOQ}\APACrefatitle {What’s in a {Smart} {Thing}? : {Development} of a
  {Multi}-layer {Taxonomy}} {What’s in a {Smart} {Thing}? : {Development} of
  a {Multi}-layer {Taxonomy}}.{\BBCQ}
\newblock
\BIn{} \APACrefbtitle {37th {International} {Conference} on {Information}
  {Systems} ({ICIS}).} {37th {International} {Conference} on {Information}
  {Systems} ({ICIS}).}
\newblock
\APACaddressPublisher{Dublin, Ireland}{AIS Electronic Library}.
\PrintBackRefs{\CurrentBib}

\bibitem [\protect \citeauthoryear {%
Raad%
\ \BBA {} Cruz%
}{%
Raad%
\ \BBA {} Cruz%
}{%
{\protect \APACyear {2015}}%
}]{%
raad2015survey}
\APACinsertmetastar {%
raad2015survey}%
\begin{APACrefauthors}%
Raad, J.%
\BCBT {}\ \BBA {} Cruz, C.%
\end{APACrefauthors}%
\unskip\
\newblock
\APACrefYearMonthDay{2015}{}{}.
\newblock
{\BBOQ}\APACrefatitle {A survey on ontology evaluation methods} {A survey on
  ontology evaluation methods}.{\BBCQ}
\newblock
\BIn{} \APACrefbtitle {Proceedings of the International Conference on Knowledge
  Engineering and Ontology Development, part of the 7th International Joint
  Conference on Knowledge Discovery, Knowledge Engineering and Knowledge
  Management.} {Proceedings of the international conference on knowledge
  engineering and ontology development, part of the 7th international joint
  conference on knowledge discovery, knowledge engineering and knowledge
  management.}
\PrintBackRefs{\CurrentBib}

\bibitem [\protect \citeauthoryear {%
Ralph%
}{%
Ralph%
}{%
{\protect \APACyear {2019}}%
}]{%
ralph_toward_2019}
\APACinsertmetastar {%
ralph_toward_2019}%
\begin{APACrefauthors}%
Ralph, P.%
\end{APACrefauthors}%
\unskip\
\newblock
\APACrefYearMonthDay{2019}{{\APACmonth{07}}}{}.
\newblock
{\BBOQ}\APACrefatitle {Toward {Methodological} {Guidelines} for {Process}
  {Theories} and {Taxonomies} in {Software} {Engineering}} {Toward
  {Methodological} {Guidelines} for {Process} {Theories} and {Taxonomies} in
  {Software} {Engineering}}.{\BBCQ}
\newblock
\APACjournalVolNumPages{IEEE Transactions on Software
  Engineering}{45}{7}{712--735}.
\PrintBackRefs{\CurrentBib}

\bibitem [\protect \citeauthoryear {%
Raza%
, Ahmad%
\BCBL {}\ \BBA {} Khan%
}{%
Raza%
\ \protect \BOthers {.}}{%
{\protect \APACyear {2018}}%
}]{%
raza_transformation_2018}
\APACinsertmetastar {%
raza_transformation_2018}%
\begin{APACrefauthors}%
Raza, U.%
, Ahmad, W.%
\BCBL {}\ \BBA {} Khan, A.%
\end{APACrefauthors}%
\unskip\
\newblock
\APACrefYearMonthDay{2018}{{\APACmonth{06}}}{}.
\newblock
{\BBOQ}\APACrefatitle {Transformation from manufacturing process taxonomy to
  repair process taxonomy: a phenetic approach} {Transformation from
  manufacturing process taxonomy to repair process taxonomy: a phenetic
  approach}.{\BBCQ}
\newblock
\APACjournalVolNumPages{Journal of Industrial Engineering
  International}{14}{2}{415--428}.
\PrintBackRefs{\CurrentBib}

\bibitem [\protect \citeauthoryear {%
Saavedra%
, Ballejos%
\BCBL {}\ \BBA {} Ale%
}{%
Saavedra%
\ \protect \BOthers {.}}{%
{\protect \APACyear {2013}}%
}]{%
saavedra_software_2013}
\APACinsertmetastar {%
saavedra_software_2013}%
\begin{APACrefauthors}%
Saavedra, R.%
, Ballejos, L\BPBI C.%
\BCBL {}\ \BBA {} Ale, M\BPBI A.%
\end{APACrefauthors}%
\unskip\
\newblock
\APACrefYearMonthDay{2013}{}{}.
\newblock
{\BBOQ}\APACrefatitle {Software Requirements Quality Evaluation: State of the
  art and research challenges} {Software requirements quality evaluation: State
  of the art and research challenges}.{\BBCQ}
\newblock
\BIn{} (\BPG~240-257).
\newblock
\APACrefnote{{ISSN}: 1850-2792}
\PrintBackRefs{\CurrentBib}

\bibitem [\protect \citeauthoryear {%
Schneider%
, Lansing%
, Gao%
\BCBL {}\ \BBA {} Sunyaev%
}{%
Schneider%
\ \protect \BOthers {.}}{%
{\protect \APACyear {2014}}%
}]{%
schneider_taxonomic_2014}
\APACinsertmetastar {%
schneider_taxonomic_2014}%
\begin{APACrefauthors}%
Schneider, S.%
, Lansing, J.%
, Gao, F.%
\BCBL {}\ \BBA {} Sunyaev, A.%
\end{APACrefauthors}%
\unskip\
\newblock
\APACrefYearMonthDay{2014}{{\APACmonth{01}}}{}.
\newblock
{\BBOQ}\APACrefatitle {A {Taxonomic} {Perspective} on {Certification}
  {Schemes}: {Development} of a {Taxonomy} for {Cloud} {Service}
  {Certification} {Criteria}} {A {Taxonomic} {Perspective} on {Certification}
  {Schemes}: {Development} of a {Taxonomy} for {Cloud} {Service}
  {Certification} {Criteria}}.{\BBCQ}
\newblock
\BIn{} \APACrefbtitle {2014 47th {Hawaii} {International} {Conference} on
  {System} {Sciences}} {2014 47th {Hawaii} {International} {Conference} on
  {System} {Sciences}}\ (\BPGS\ 4998--5007).
\newblock
\APACaddressPublisher{Waikoloa, HI, USA}{IEEE}.
\PrintBackRefs{\CurrentBib}

\bibitem [\protect \citeauthoryear {%
Schäffer%
\ \BBA {} Stelzer%
}{%
Schäffer%
\ \BBA {} Stelzer%
}{%
{\protect \APACyear {2017}}%
}]{%
schaffer_assessing_2017}
\APACinsertmetastar {%
schaffer_assessing_2017}%
\begin{APACrefauthors}%
Schäffer, T.%
\BCBT {}\ \BBA {} Stelzer, D.%
\end{APACrefauthors}%
\unskip\
\newblock
\APACrefYearMonthDay{2017}{{\APACmonth{02}}}{}.
\newblock
{\BBOQ}\APACrefatitle {Assessing {Tools} for {Coordinating} {Quality} of
  {Master} {Data} in {Inter}-organizational {Product} {Information} {Sharing}}
  {Assessing {Tools} for {Coordinating} {Quality} of {Master} {Data} in
  {Inter}-organizational {Product} {Information} {Sharing}}.{\BBCQ}
\newblock
\BIn{} \APACrefbtitle {13. {Internationalen} {Tagung} {Wirtschaftsinformatik}.}
  {13. {Internationalen} {Tagung} {Wirtschaftsinformatik}.}
\newblock
\APACaddressPublisher{St. Gallen, Switzerland}{}.
\PrintBackRefs{\CurrentBib}

\bibitem [\protect \citeauthoryear {%
Seyffarth%
, Kühnel%
\BCBL {}\ \BBA {} Sackmann%
}{%
Seyffarth%
\ \protect \BOthers {.}}{%
{\protect \APACyear {2017}}%
}]{%
seyffarth_taxonomy_2017}
\APACinsertmetastar {%
seyffarth_taxonomy_2017}%
\begin{APACrefauthors}%
Seyffarth, T.%
, Kühnel, S.%
\BCBL {}\ \BBA {} Sackmann, S.%
\end{APACrefauthors}%
\unskip\
\newblock
\APACrefYearMonthDay{2017}{}{}.
\newblock
{\BBOQ}\APACrefatitle {A {Taxonomy} of {Compliance} {Processes} for {Business}
  {Process} {Compliance}} {A {Taxonomy} of {Compliance} {Processes} for
  {Business} {Process} {Compliance}}.{\BBCQ}
\newblock
\BIn{} \APACrefbtitle {Proceedings {International} {Conference} on {Business}
  {Process} {Management}} {Proceedings {International} {Conference} on
  {Business} {Process} {Management}}\ (\BPGS\ 71--87).
\newblock
\APACaddressPublisher{Barcelona, Spain}{Springer International Publishing}.
\PrintBackRefs{\CurrentBib}

\bibitem [\protect \citeauthoryear {%
Siering%
, Clapham%
, Engel%
\BCBL {}\ \BBA {} Gomber%
}{%
Siering%
\ \protect \BOthers {.}}{%
{\protect \APACyear {2017}}%
}]{%
siering_taxonomy_2017}
\APACinsertmetastar {%
siering_taxonomy_2017}%
\begin{APACrefauthors}%
Siering, M.%
, Clapham, B.%
, Engel, O.%
\BCBL {}\ \BBA {} Gomber, P.%
\end{APACrefauthors}%
\unskip\
\newblock
\APACrefYearMonthDay{2017}{{\APACmonth{09}}}{}.
\newblock
{\BBOQ}\APACrefatitle {A taxonomy of financial market manipulations:
  establishing trust and market integrity in the financialized economy through
  automated fraud detection} {A taxonomy of financial market manipulations:
  establishing trust and market integrity in the financialized economy through
  automated fraud detection}.{\BBCQ}
\newblock
\APACjournalVolNumPages{Journal of Information Technology}{32}{3}{251--269}.
\PrintBackRefs{\CurrentBib}

\bibitem [\protect \citeauthoryear {%
Snow%
\ \BBA {} Reck%
}{%
Snow%
\ \BBA {} Reck%
}{%
{\protect \APACyear {2016}}%
}]{%
snow_developing_2016}
\APACinsertmetastar {%
snow_developing_2016}%
\begin{APACrefauthors}%
Snow, N\BPBI M.%
\BCBT {}\ \BBA {} Reck, J\BPBI L.%
\end{APACrefauthors}%
\unskip\
\newblock
\APACrefYearMonthDay{2016}{{\APACmonth{01}}}{}.
\newblock
{\BBOQ}\APACrefatitle {Developing a {Government} {Reporting} {Taxonomy}}
  {Developing a {Government} {Reporting} {Taxonomy}}.{\BBCQ}
\newblock
\APACjournalVolNumPages{Journal of Information Systems}{30}{2}{49--81}.
\PrintBackRefs{\CurrentBib}

\bibitem [\protect \citeauthoryear {%
Strasser%
}{%
Strasser%
}{%
{\protect \APACyear {2017}}%
}]{%
strasser_delphi_2017}
\APACinsertmetastar {%
strasser_delphi_2017}%
\begin{APACrefauthors}%
Strasser, A.%
\end{APACrefauthors}%
\unskip\
\newblock
\APACrefYearMonthDay{2017}{{\APACmonth{10}}}{}.
\newblock
{\BBOQ}\APACrefatitle {Delphi {Method} {Variants} in {Information} {Systems}
  {Research}: {Taxonomy} {Development} and {Application}} {Delphi {Method}
  {Variants} in {Information} {Systems} {Research}: {Taxonomy} {Development}
  and {Application}}.{\BBCQ}
\newblock
\APACjournalVolNumPages{Electronic Journal of Business Research
  Methods}{15}{}{}.
\PrintBackRefs{\CurrentBib}

\bibitem [\protect \citeauthoryear {%
Strode%
}{%
Strode%
}{%
{\protect \APACyear {2016}}%
}]{%
strode_dependency_2016}
\APACinsertmetastar {%
strode_dependency_2016}%
\begin{APACrefauthors}%
Strode, D\BPBI E.%
\end{APACrefauthors}%
\unskip\
\newblock
\APACrefYearMonthDay{2016}{{\APACmonth{02}}}{}.
\newblock
{\BBOQ}\APACrefatitle {A dependency taxonomy for agile software development
  projects} {A dependency taxonomy for agile software development
  projects}.{\BBCQ}
\newblock
\APACjournalVolNumPages{Information Systems Frontiers}{18}{1}{23--46}.
\PrintBackRefs{\CurrentBib}

\bibitem [\protect \citeauthoryear {%
Stöckli%
, Uebernickel%
\BCBL {}\ \BBA {} Brenner%
}{%
Stöckli%
\ \protect \BOthers {.}}{%
{\protect \APACyear {2017}}%
}]{%
stockli_capturing_2017}
\APACinsertmetastar {%
stockli_capturing_2017}%
\begin{APACrefauthors}%
Stöckli, E.%
, Uebernickel, F.%
\BCBL {}\ \BBA {} Brenner, W.%
\end{APACrefauthors}%
\unskip\
\newblock
\APACrefYearMonthDay{2017}{{\APACmonth{08}}}{}.
\newblock
{\BBOQ}\APACrefatitle {Capturing {Functional} {Affordances} of {Enterprise}
  {Social} {Software}} {Capturing {Functional} {Affordances} of {Enterprise}
  {Social} {Software}}.{\BBCQ}
\newblock
\BIn{} \APACrefbtitle {Proceedings of the 23rd {Americas} {Conference} on
  {Information} {Systems}.} {Proceedings of the 23rd {Americas} {Conference} on
  {Information} {Systems}.}
\newblock
\APACaddressPublisher{Boston, MA, USA}{AIS Electronic Library}.
\PrintBackRefs{\CurrentBib}

\bibitem [\protect \citeauthoryear {%
Szopinski%
, Schoormann%
\BCBL {}\ \BBA {} Kundisch%
}{%
Szopinski%
\ \protect \BOthers {.}}{%
{\protect \APACyear {2020}}%
}]{%
szopinski_criteria_2020}
\APACinsertmetastar {%
szopinski_criteria_2020}%
\begin{APACrefauthors}%
Szopinski, D.%
, Schoormann, T.%
\BCBL {}\ \BBA {} Kundisch, D.%
\end{APACrefauthors}%
\unskip\
\newblock
\APACrefYearMonthDay{2020}{{\APACmonth{01}}}{}.
\newblock
{\BBOQ}\APACrefatitle {Criteria as a {Prelude} for {Guiding} {Taxonomy}
  {Evaluation}} {Criteria as a {Prelude} for {Guiding} {Taxonomy}
  {Evaluation}}.{\BBCQ}
\newblock
\BIn{} \APACrefbtitle {Proceedings of the 53rd {Hawai}’i {International}
  {Conference} on {System} {Sciences} ({HICSS})} {Proceedings of the 53rd
  {Hawai}’i {International} {Conference} on {System} {Sciences} ({HICSS})}\
  (\BPGS\ 1--10).
\newblock
\APACaddressPublisher{Hawaii, USA}{ScholarSpace}.
\PrintBackRefs{\CurrentBib}

\bibitem [\protect \citeauthoryear {%
Thiebes%
, Kleiber%
\BCBL {}\ \BBA {} Sunyaev%
}{%
Thiebes%
\ \protect \BOthers {.}}{%
{\protect \APACyear {2017}}%
}]{%
thiebes_cancer_2017}
\APACinsertmetastar {%
thiebes_cancer_2017}%
\begin{APACrefauthors}%
Thiebes, S.%
, Kleiber, G.%
\BCBL {}\ \BBA {} Sunyaev, A.%
\end{APACrefauthors}%
\unskip\
\newblock
\APACrefYearMonthDay{2017}{}{}.
\newblock
{\BBOQ}\APACrefatitle {Cancer {Genomics} {Research} in the {Cloud}: {A}
  {Taxonomy} of {Genome} {Data} {Sets}} {Cancer {Genomics} {Research} in the
  {Cloud}: {A} {Taxonomy} of {Genome} {Data} {Sets}}.{\BBCQ}
\newblock
\BIn{} \APACrefbtitle {4th {International} {Workshop} on {Genome} {Privacy} and
  {Security}.} {4th {International} {Workshop} on {Genome} {Privacy} and
  {Security}.}
\newblock
\APACaddressPublisher{Orlando, Florida, USA}{}.
\PrintBackRefs{\CurrentBib}

\bibitem [\protect \citeauthoryear {%
Tilly%
, Posegga%
, Fischbach%
\BCBL {}\ \BBA {} Schoder%
}{%
Tilly%
\ \protect \BOthers {.}}{%
{\protect \APACyear {2017}}%
}]{%
tilly_towards_2017}
\APACinsertmetastar {%
tilly_towards_2017}%
\begin{APACrefauthors}%
Tilly, R.%
, Posegga, O.%
, Fischbach, K.%
\BCBL {}\ \BBA {} Schoder, D.%
\end{APACrefauthors}%
\unskip\
\newblock
\APACrefYearMonthDay{2017}{}{}.
\newblock
{\BBOQ}\APACrefatitle {Towards a {Conceptualization} of {Data} and
  {Information} {Quality} in {Social} {Information} {Systems}} {Towards a
  {Conceptualization} of {Data} and {Information} {Quality} in {Social}
  {Information} {Systems}}.{\BBCQ}
\newblock
\APACjournalVolNumPages{Business \& Information Systems
  Engineering}{59}{1}{3--21}.
\PrintBackRefs{\CurrentBib}

\bibitem [\protect \citeauthoryear {%
Tom%
, Aurum%
\BCBL {}\ \BBA {} Vidgen%
}{%
Tom%
\ \protect \BOthers {.}}{%
{\protect \APACyear {2013}}%
}]{%
tom_exploration_2013}
\APACinsertmetastar {%
tom_exploration_2013}%
\begin{APACrefauthors}%
Tom, E.%
, Aurum, A.%
\BCBL {}\ \BBA {} Vidgen, R.%
\end{APACrefauthors}%
\unskip\
\newblock
\APACrefYearMonthDay{2013}{{\APACmonth{06}}}{}.
\newblock
{\BBOQ}\APACrefatitle {An exploration of technical debt} {An exploration of
  technical debt}.{\BBCQ}
\newblock
\APACjournalVolNumPages{Journal of Systems and Software}{86}{6}{1498--1516}.
\PrintBackRefs{\CurrentBib}

\bibitem [\protect \citeauthoryear {%
Tsatsou%
, Elaluf-Calderwood%
\BCBL {}\ \BBA {} Liebenau%
}{%
Tsatsou%
\ \protect \BOthers {.}}{%
{\protect \APACyear {2010}}%
}]{%
tsatsou_towards_2010}
\APACinsertmetastar {%
tsatsou_towards_2010}%
\begin{APACrefauthors}%
Tsatsou, P.%
, Elaluf-Calderwood, S.%
\BCBL {}\ \BBA {} Liebenau, J.%
\end{APACrefauthors}%
\unskip\
\newblock
\APACrefYearMonthDay{2010}{{\APACmonth{09}}}{}.
\newblock
{\BBOQ}\APACrefatitle {Towards a taxonomy for regulatory issues in a digital
  business ecosystem in the {EU}} {Towards a taxonomy for regulatory issues in
  a digital business ecosystem in the {EU}}.{\BBCQ}
\newblock
\APACjournalVolNumPages{Journal of Information Technology}{25}{3}{288--307}.
\PrintBackRefs{\CurrentBib}

\bibitem [\protect \citeauthoryear {%
Tönnissen%
\ \BBA {} Teuteberg%
}{%
Tönnissen%
\ \BBA {} Teuteberg%
}{%
{\protect \APACyear {2018}}%
}]{%
tonnissen_towards_2018}
\APACinsertmetastar {%
tonnissen_towards_2018}%
\begin{APACrefauthors}%
Tönnissen, S.%
\BCBT {}\ \BBA {} Teuteberg, F.%
\end{APACrefauthors}%
\unskip\
\newblock
\APACrefYearMonthDay{2018}{{\APACmonth{06}}}{}.
\newblock
{\BBOQ}\APACrefatitle {Towards a {Taxonomy} for {Smart} {Contracts}} {Towards a
  {Taxonomy} for {Smart} {Contracts}}.{\BBCQ}
\newblock
\BIn{} \APACrefbtitle {Proceedings {European} {Conference} on {Information}
  {Systems}.} {Proceedings {European} {Conference} on {Information} {Systems}.}
\newblock
\APACaddressPublisher{Portsmouth, UK}{AIS Electronic Library}.
\PrintBackRefs{\CurrentBib}

\bibitem [\protect \citeauthoryear {%
Unterkalmsteiner%
}{%
Unterkalmsteiner%
}{%
{\protect \APACyear {2020}}%
}]{%
unterkalmsteiner_early_2020}
\APACinsertmetastar {%
unterkalmsteiner_early_2020}%
\begin{APACrefauthors}%
Unterkalmsteiner, M.%
\end{APACrefauthors}%
\unskip\
\newblock
\APACrefYearMonthDay{2020}{}{}.
\newblock
{\BBOQ}\APACrefatitle {Early {Requirements} {Traceability} with
  {Domain}-{Specific} {Taxonomies} - {A} {Pilot} {Experiment}} {Early
  {Requirements} {Traceability} with {Domain}-{Specific} {Taxonomies} - {A}
  {Pilot} {Experiment}}.{\BBCQ}
\newblock
\BIn{} \APACrefbtitle {28th {IEEE} {International} {Requirements} {Engineering}
  {Conference}} {28th {IEEE} {International} {Requirements} {Engineering}
  {Conference}}\ (\BPGS\ 323--327).
\newblock
\APACaddressPublisher{Zurich, Switzerland}{IEEE}.
\PrintBackRefs{\CurrentBib}

\bibitem [\protect \citeauthoryear {%
Unterkalmsteiner%
\ \BBA {} Abdeen%
}{%
Unterkalmsteiner%
\ \BBA {} Abdeen%
}{%
{\protect \APACyear {2021}}%
}]{%
unterkalmsteiner_supp_2021}
\APACinsertmetastar {%
unterkalmsteiner_supp_2021}%
\begin{APACrefauthors}%
Unterkalmsteiner, M.%
\BCBT {}\ \BBA {} Abdeen, W.%
\end{APACrefauthors}%
\unskip\
\newblock
\APACrefYearMonthDay{2021}{}{}.
\newblock
\APACrefbtitle {{Supplementary material for: A compendium and evaluation of
  taxonomy quality attributes}.} {{Supplementary material for: A compendium and
  evaluation of taxonomy quality attributes}.}
\newblock
\APACaddressPublisher{}{Zenodo}.
\newblock
\begin{APACrefURL} \url{https://doi.org/10.5281/zenodo.5148475}
  \end{APACrefURL}
\newblock
\begin{APACrefDOI} 10.5281/zenodo.5148475 \end{APACrefDOI}
\PrintBackRefs{\CurrentBib}

\bibitem [\protect \citeauthoryear {%
Usman%
, Britto%
, Börstler%
\BCBL {}\ \BBA {} Mendes%
}{%
Usman%
\ \protect \BOthers {.}}{%
{\protect \APACyear {2017}}%
}]{%
usman_taxonomies_2017}
\APACinsertmetastar {%
usman_taxonomies_2017}%
\begin{APACrefauthors}%
Usman, M.%
, Britto, R.%
, Börstler, J.%
\BCBL {}\ \BBA {} Mendes, E.%
\end{APACrefauthors}%
\unskip\
\newblock
\APACrefYearMonthDay{2017}{{\APACmonth{05}}}{}.
\newblock
{\BBOQ}\APACrefatitle {Taxonomies in software engineering: {A} {Systematic}
  mapping study and a revised taxonomy development method} {Taxonomies in
  software engineering: {A} {Systematic} mapping study and a revised taxonomy
  development method}.{\BBCQ}
\newblock
\APACjournalVolNumPages{Information and Software Technology}{85}{}{43--59}.
\PrintBackRefs{\CurrentBib}

\bibitem [\protect \citeauthoryear {%
Wallace%
\ \BBA {} Ross%
}{%
Wallace%
\ \BBA {} Ross%
}{%
{\protect \APACyear {2006}}%
}]{%
wallace_beyond_2006}
\APACinsertmetastar {%
wallace_beyond_2006}%
\begin{APACrefauthors}%
Wallace, B.%
\BCBT {}\ \BBA {} Ross, A.%
\end{APACrefauthors}%
\unskip\
\newblock
\APACrefYear{2006}.
\newblock
\APACrefbtitle {Beyond {Human} {Error}: {Taxonomies} and {Safety} {Science}}
  {Beyond {Human} {Error}: {Taxonomies} and {Safety} {Science}}\
  (\PrintOrdinal{1st edition}\ \BEd).
\newblock
\APACaddressPublisher{Boca Raton, FL}{CRC Press}.
\PrintBackRefs{\CurrentBib}

\bibitem [\protect \citeauthoryear {%
Werder%
\ \BBA {} Wang%
}{%
Werder%
\ \BBA {} Wang%
}{%
{\protect \APACyear {2016}}%
}]{%
werder_towards_2016}
\APACinsertmetastar {%
werder_towards_2016}%
\begin{APACrefauthors}%
Werder, K.%
\BCBT {}\ \BBA {} Wang, H\BHBI Y.%
\end{APACrefauthors}%
\unskip\
\newblock
\APACrefYearMonthDay{2016}{}{}.
\newblock
{\BBOQ}\APACrefatitle {Towards a {Software} {Product} {Industry}
  {Classification}} {Towards a {Software} {Product} {Industry}
  {Classification}}.{\BBCQ}
\newblock
\APACjournalVolNumPages{New Trends in Software Methodologies, Tools and
  Techniques}{}{}{27--37}.
\PrintBackRefs{\CurrentBib}

\bibitem [\protect \citeauthoryear {%
Williams%
, Chatterjee%
\BCBL {}\ \BBA {} Rossi%
}{%
Williams%
\ \protect \BOthers {.}}{%
{\protect \APACyear {2008}}%
}]{%
williams_design_2008}
\APACinsertmetastar {%
williams_design_2008}%
\begin{APACrefauthors}%
Williams, K.%
, Chatterjee, S.%
\BCBL {}\ \BBA {} Rossi, M.%
\end{APACrefauthors}%
\unskip\
\newblock
\APACrefYearMonthDay{2008}{{\APACmonth{10}}}{}.
\newblock
{\BBOQ}\APACrefatitle {Design of emerging digital services: a taxonomy} {Design
  of emerging digital services: a taxonomy}.{\BBCQ}
\newblock
\APACjournalVolNumPages{European Journal of Information
  Systems}{17}{5}{505--517}.
\PrintBackRefs{\CurrentBib}

\bibitem [\protect \citeauthoryear {%
{World Health Organization}%
}{%
{World Health Organization}%
}{%
{\protect \APACyear {2021}}%
}]{%
world_health_organization_classification_2021}
\APACinsertmetastar {%
world_health_organization_classification_2021}%
\begin{APACrefauthors}%
{World Health Organization}.%
\end{APACrefauthors}%
\unskip\
\newblock
\APACrefYearMonthDay{2021}{}{}.
\newblock
\APACrefbtitle {Classification of {Diseases} ({ICD}).} {Classification of
  {Diseases} ({ICD}).}
\newblock
\begin{APACrefURL}
  [{2021-05-15}]\url{https://www.who.int/standards/classifications/classification-of-diseases}
  \end{APACrefURL}
\PrintBackRefs{\CurrentBib}

\bibitem [\protect \citeauthoryear {%
Zrenner%
, Hassan%
, Otto%
\BCBL {}\ \BBA {} Marx~Gómez%
}{%
Zrenner%
\ \protect \BOthers {.}}{%
{\protect \APACyear {2017}}%
}]{%
zrenner_data_2017}
\APACinsertmetastar {%
zrenner_data_2017}%
\begin{APACrefauthors}%
Zrenner, J.%
, Hassan, A\BPBI P.%
, Otto, B.%
\BCBL {}\ \BBA {} Marx~Gómez, J\BPBI C.%
\end{APACrefauthors}%
\unskip\
\newblock
\APACrefYearMonthDay{2017}{}{}.
\newblock
{\BBOQ}\APACrefatitle {Data source taxonomy for supply network structure
  visibility} {Data source taxonomy for supply network structure
  visibility}.{\BBCQ}
\newblock
\BIn{} \APACrefbtitle {Proceedings of the {Hamburg} {International}
  {Conference} of {Logistics} ({HICL})} {Proceedings of the {Hamburg}
  {International} {Conference} of {Logistics} ({HICL})}\ (\BPGS\ 117--137).
\newblock
\APACaddressPublisher{Hamburg, Germany}{epubli}.
\PrintBackRefs{\CurrentBib}

\end{thebibliography}

\end{document}